\documentclass[preprint]{ptephy_v1}

\newcommand{\vecP}{\mbox{\boldmath $ P $}}
\newcommand{\vecW}{\mbox{\boldmath $ W $}}
\newcommand{\vecX}{\mbox{\boldmath $ X $}}
\newcommand{\vecZ}{\mbox{\boldmath $ Z $}}
\newcommand{\vecLambda}{\mbox{\boldmath $ \Lambda $}}

\begin{document}

\title{Decoherence in Sub-Systems of an Isolated System and the Disappearance of Quantum Multiverse}


\author{Takuji ISHIKAWA}





\begin{abstract}%
Using three particles model without external environments, it is found that decisions of respective state of three particles by decoherence are not simultaneous. Furthermore, in this model, wave function of total three body system collapses spontaneously without any external environments. Therefore  we may able to insist that a wavefunction of our universe has already collapsed spontaneously without any external observer, because of the same mechanism with this model.
\end{abstract}

\subjectindex{xxxx, xxx}

\maketitle

\section{Introduction}
Decoherence is that interference among different quantum states disappear because of dissipation-fluctuation from environments[1,2]. It is said that decoherence makes a quantum system classical.  Because when interference vanishes, an object can not go to other quantum states. It means the object gets classicality. In quantum mechanics, an object can be in some contradictory states at the same time. This fact has made some paradoxes. Now, it is expected that decoherence solves those paradoxes.  For example, "Schr\"odinger's cat", which is a superposition of a dead cat and an alive cat, disappears by decoherence because of environmental effect such as room temperature. Therefore we can not see real Schr\"odinger's cat. On the other hand, decoherence seems helpless for fantasies such as Everett's interpretation (Many-worlds interpretation)[3] or multiverse. In principle, the universe we live in can be in some contradictory states at the same time. But there is no observer who decides a state of our universe uniquely nor external environment making decoherence for our universe. So, do we have better lives in the other universe ?

 I have studied decoherence in a finite system. My motivation was the fact that semi-classical models, such as liquid drop model[7] or TDHF[8], are not bad theories in nuclear physics. I wondered the fact. Then I knew Caldeira-Leggett's theory[5,6]. It is said that the key of decoherence is the external environment which is composed with an infinite number of degrees of freedom. Although I was very impressed, I doubted the "infinite numbers". Because I thought that nuclear system, which is finite system, is classical to some extent.

 In my previous paper[4], I have shown that decoherence occurs in a sub-system of an isolated system  without any other external environments. In the paper[4], there are only 3 particles, and it is suggested that particle-1 as a sub-system gets classicality because of fluctuations from other 2 particles. Furthermore, it is suggested that while classical trajectories are crossing, decoherences arise in corresponding quantum system.

 Because my model is symmetric about three particles, we can confirm that the environmental particles, particle-2 and -3 also get classicality respectively. Then, how are the 3 particles as a whole?

 By this research, it is suggested that a state of a whole system of three particles seems to get classicality too. That is, this system's wavefunction collapses spontaneously without any external environments. This result seems to break unitarity of this whole system of three particles. 

\section{Method}

Please imagine a closed system, there three particles are mutually tied with springs which have different angular frequencies respectively. Lagrangean \(L\) is as follow. (\(x_1\) means position of particle-1, etc.)
{\small
\begin{eqnarray}
 L &=& \frac{m}{2} \dot{x_1}^{2} +  \frac{m}{2} \dot{x_2}^{2} + \frac{m}{2} \dot{x_3}^{2} \nonumber \\
 &\quad& -\frac{m}{2} \omega_{12}^{2} (x_1-x_2)^{2} -\frac{m}{2} \omega_{13}^{2} (x_1-x_3)^{2} \nonumber \\
 &\quad& \quad -\frac{m}{2} \omega_{23}^{2} (x_2-x_3)^{2}  \label{e1}
\end{eqnarray}
}
  Here, there are three angular frequencies, \( \omega_{12} \), \( \omega_{13} \), \( \omega_{23} \). When they do not have the ratio of whole number, their trajectories are not closed, and densely fill some finite region. There, since it is thought that ergodicity is fulfilled locally, it can be expected that statistical property and also irreversibility appear. If irreversibility induces decoherence, decoherence should arise by such a system.  For this model, I will apply the technique of Caldeira-Leggett. This three particles model is a extreme reduction of their ''harmonic oscillator plus reservoir model''[5,6].  We can derive a Feynman propagator for Eq.(\( \ref{e1}  \)). Using the propagator, we can write a time evolution of wave function of three body system.

Initial wave function of the three body system is the product of wave functions of each particles at initial time $t_0$. And the each initial state is Schr\"odinger cat state,
{\small
\begin{eqnarray}
 &&\psi_1(x_{1(0)},t_0) = \tilde{ N_1 } \times \quad\quad \quad \nonumber \\
 \nonumber \\
 &&\left[ \  \exp\left\{-\frac{x_{1(0)}^2}{4\sigma_1^2}\right\}+\exp\left\{ -\frac{(x_{1(0)}-d_1)^2}{4\sigma_1^2}\right\} \ \right] \nonumber \\ \label{e3} 
\end{eqnarray}
}
etc.. Here, \(x_{1(0)}\) means \(x_1(t_0)\), \(\sigma_1\) means half width of packet and \(\tilde{ N_1 }\) means a normalization constant. When we are only interested in the information about a degree of freedom (particle-1) as a subsystem,  we should integrate out the information about particle-2 and -3 as environments. Then we can get the information about particle-1 only, that is, the reduced density function for particle-1, $\tilde{\rho}_1$.
{\small
\begin{eqnarray}
 &&\tilde{\rho}_1^{(reduced)}(x_1,t) = \nonumber \\ \nonumber \\
 &&\quad\quad\quad \int_{-\infty}^{\infty} \int_{-\infty}^{\infty}dx_2 \ dx_3 \ \rho^{(total)}(x_1,x_2,x_3,t) \nonumber \\ \label{e4}
\end{eqnarray}
}

 Furthermore, we can separate the quantum interference term from the reduced density, and a disappearance of the interference means decoherence. I used numerical integration in Eq.(\ref{e4}) and final normalization. If you interested in details, please watch Appendix. 
 We can get the reduced density function for particle-2 and particle-3 respectively with same ways. Therefore we can also check decoherence for particle-2 and particle-3.

 This procedure above is basically the same as Caldeira-Leggett's technique[5,6]. They used an influence functional method in which a Feynman propagator includes effects of environmental degrees of freedom. The difference between my procedure here and theirs is only an order of integrals and path integrals.

 On the other hand, we can draw corresponding classical trajectories \(x_1(t),x_2(t),x_3(t)\). Each particle has 2 initial positions which are corresponding to centers of two Gaussian packets in quantum system, and their all initial velocities are set 0. Then we can draw $2^3 = 8$ trajectories on a ($x_i$-t) plane (\(i=1,2,3\)).  While classical trajectories are crossing, decoherences arise in corresponding quantum system[4].

\section{Result}

\begin{figure}[!]

\begin{center}

\begin{minipage}{.45\linewidth}
(a) t=3.005\\
\includegraphics[width=\linewidth]{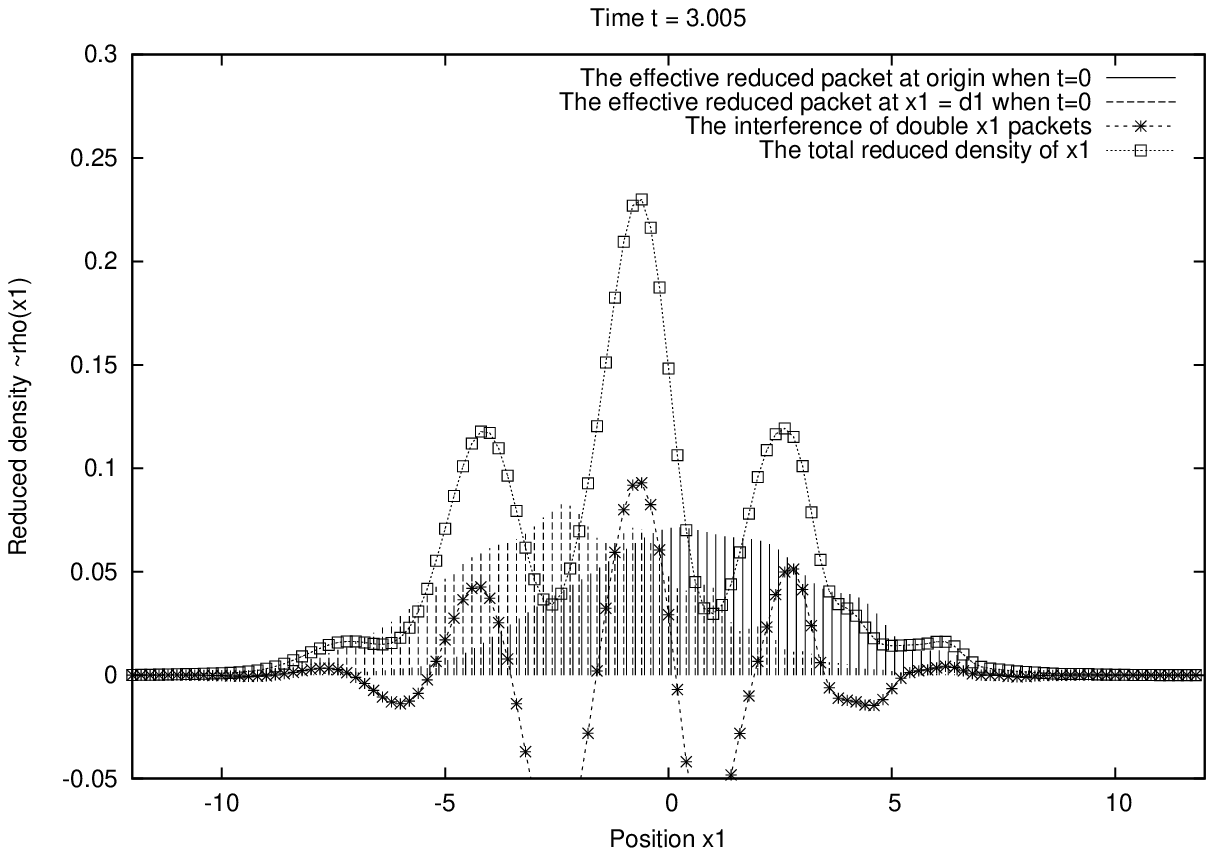}
\end{minipage}

\begin{minipage}{.45\linewidth}
(b) t=3.505\\
\includegraphics[width=\linewidth]{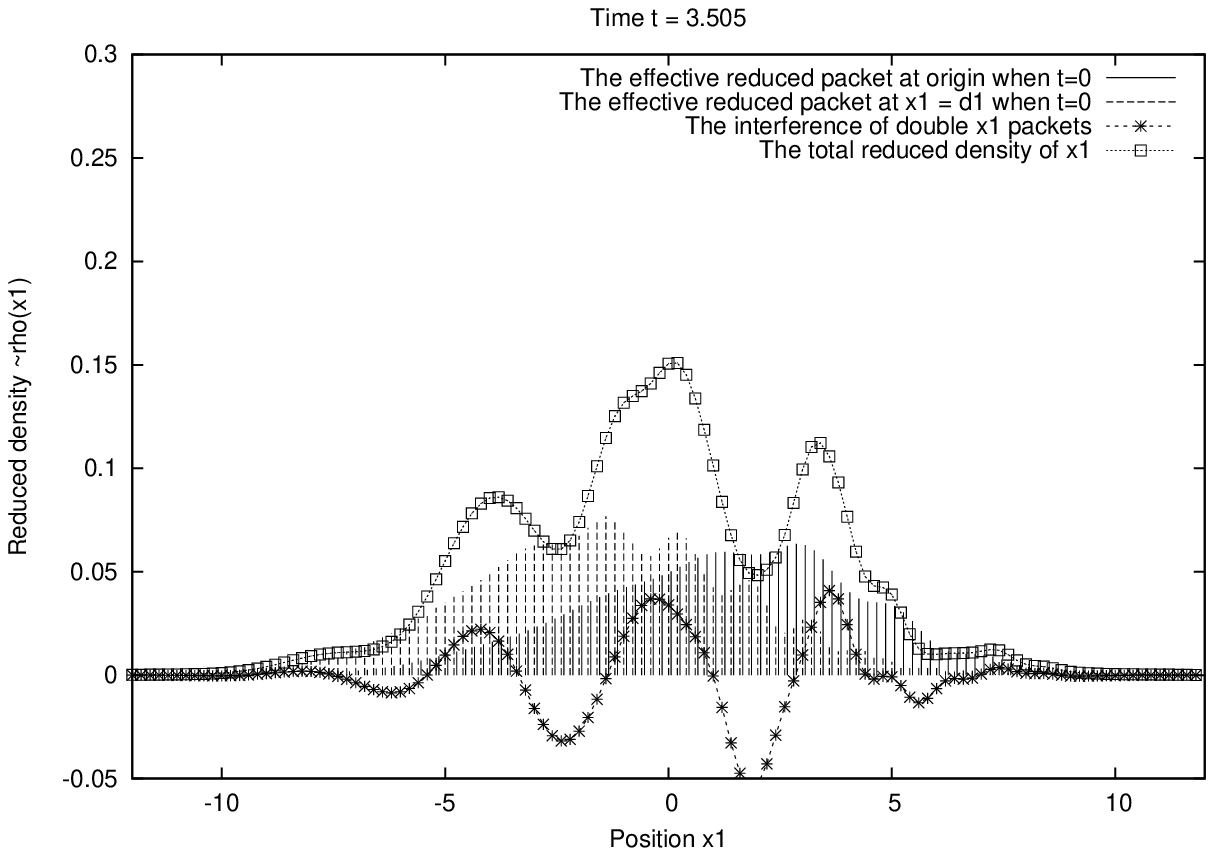}
\end{minipage}

\begin{minipage}{.45\linewidth}
(c) t=6.005\\
\includegraphics[width=\linewidth]{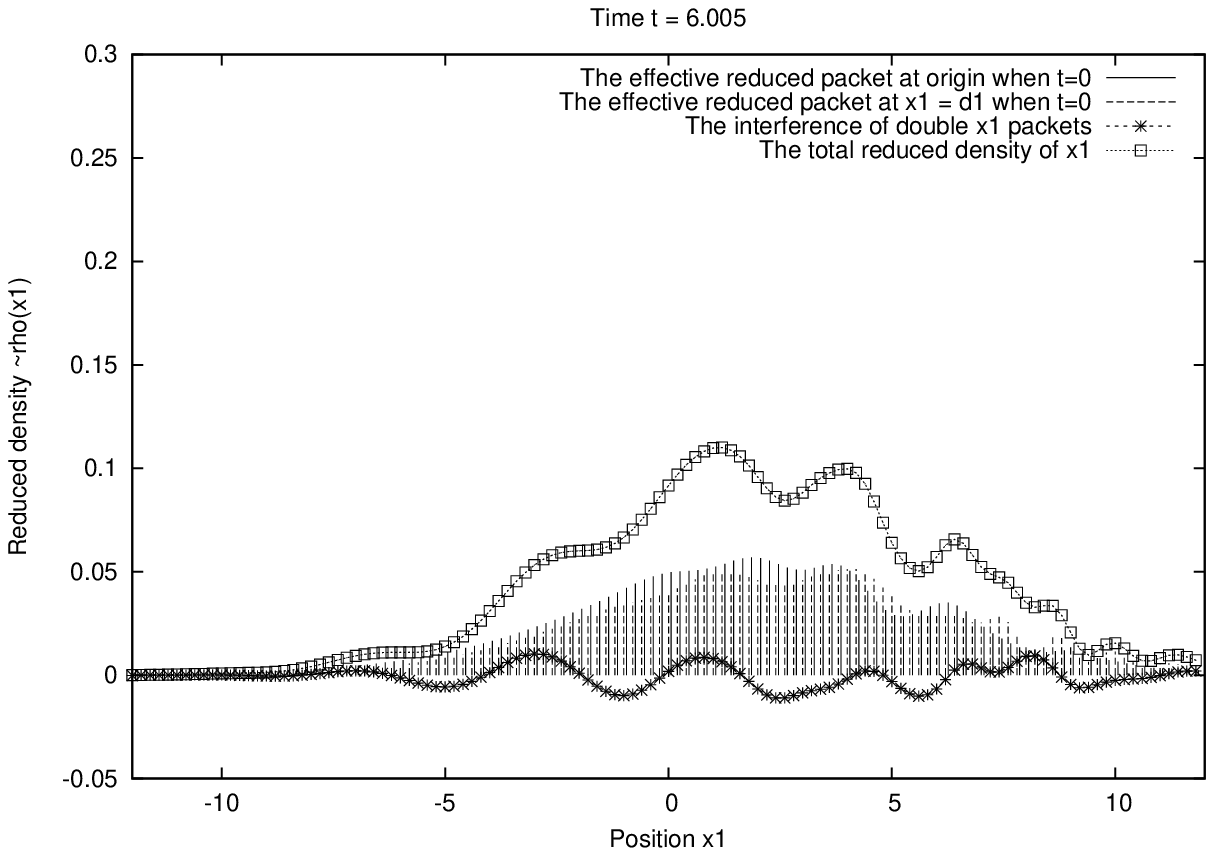}
\end{minipage}

\begin{minipage}{.5\linewidth}
(d) Classical trajectories \(x_1\)(t)\\
\includegraphics[width=\linewidth]{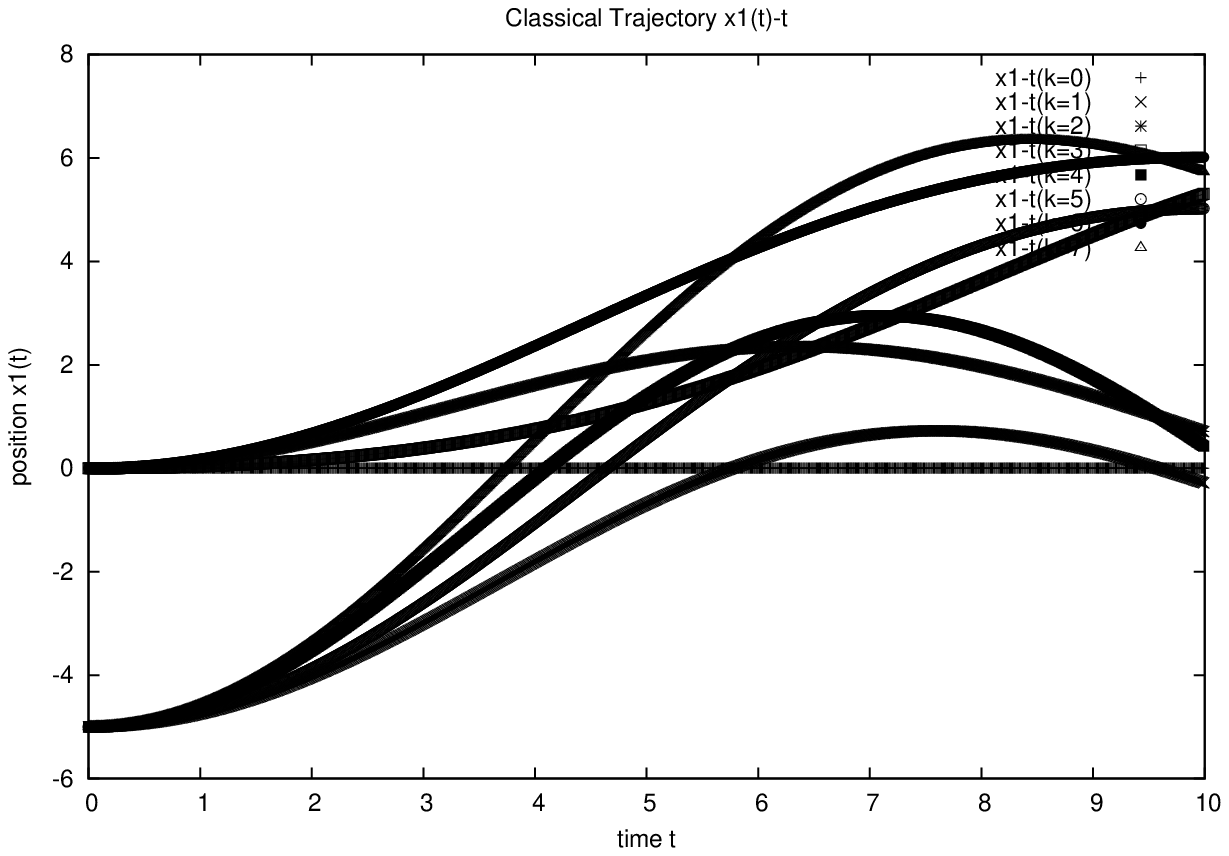}
\end{minipage}

\end{center}

\caption{ Time evolution of the reduced dencity of particle-1. (a) Time t=3.005, when interference is still strong. (b) Time t=3.505, when the interference starts to damp. (c) Time t=6.005, when decoherence arises. (d) Classical trajectories of particle-1, position \(x_1\)(t) vs time t. Trajectories cross after time t=3.505.}
\label{f1}
\end{figure}

\begin{figure}[!]

\begin{center}

\begin{minipage}{.45\linewidth}
(a) t=4.005\\
\includegraphics[width=\linewidth]{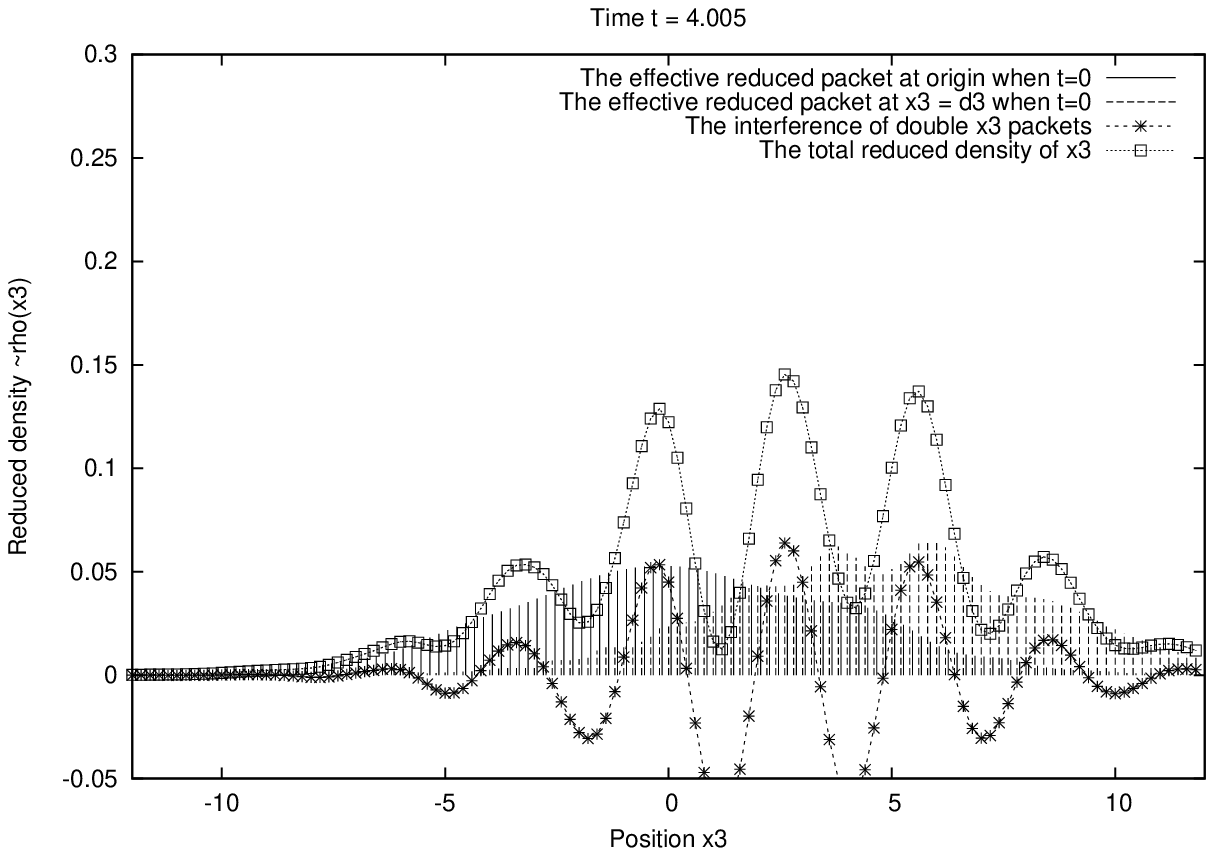}
\end{minipage}

\begin{minipage}{.45\linewidth}
(b) t=5.505\\
\includegraphics[width=\linewidth]{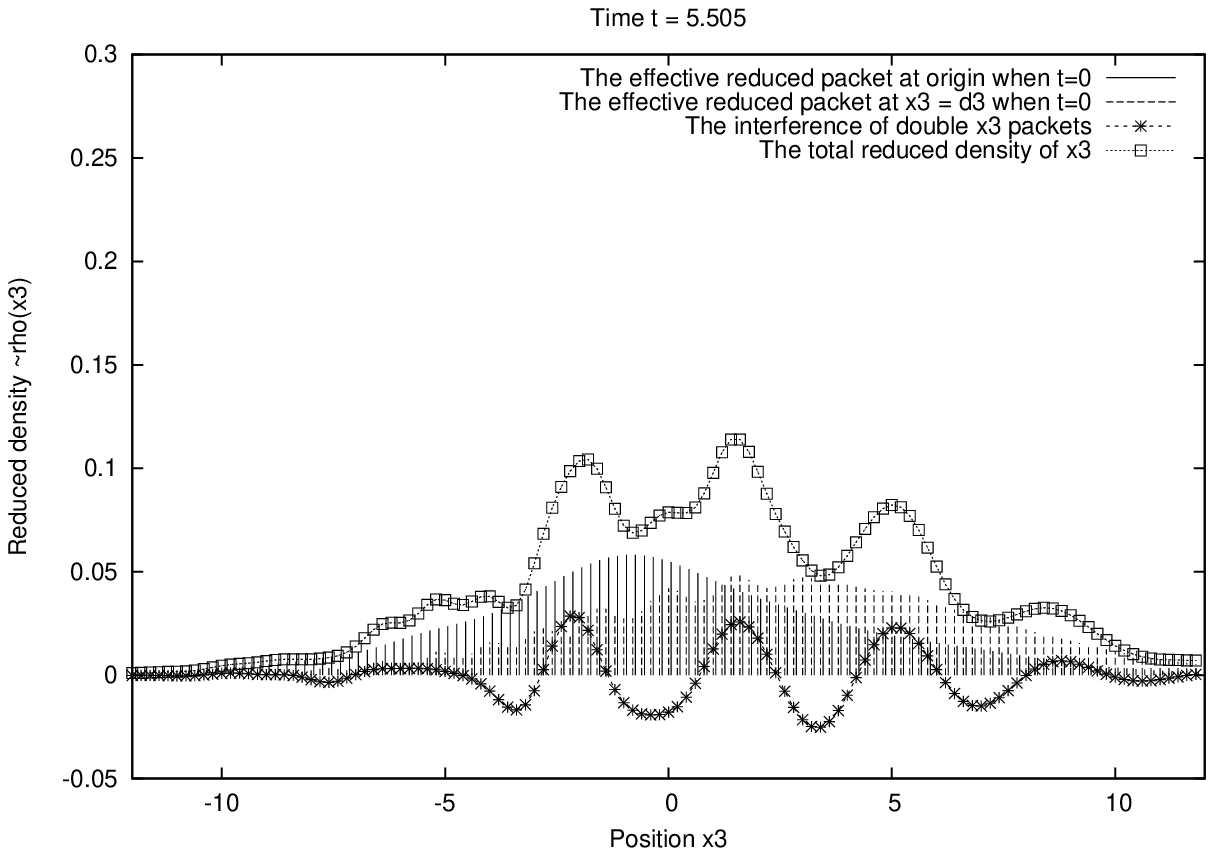}
\end{minipage}

\begin{minipage}{.45\linewidth}
(c) t=10.005\\
\includegraphics[width=\linewidth]{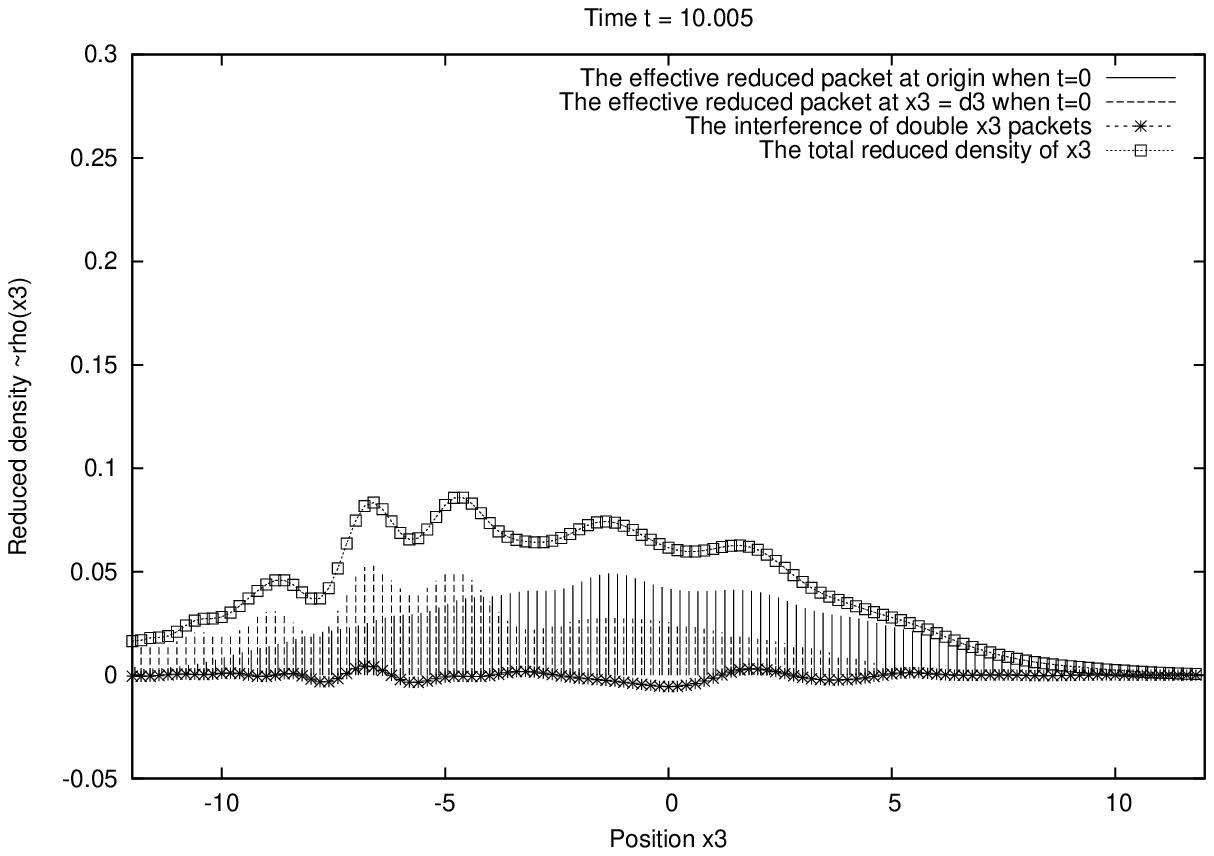}
\end{minipage}

\begin{minipage}{.5\linewidth}
(d) Classical trajectories \(x_3\)(t)\\
\includegraphics[width=\linewidth]{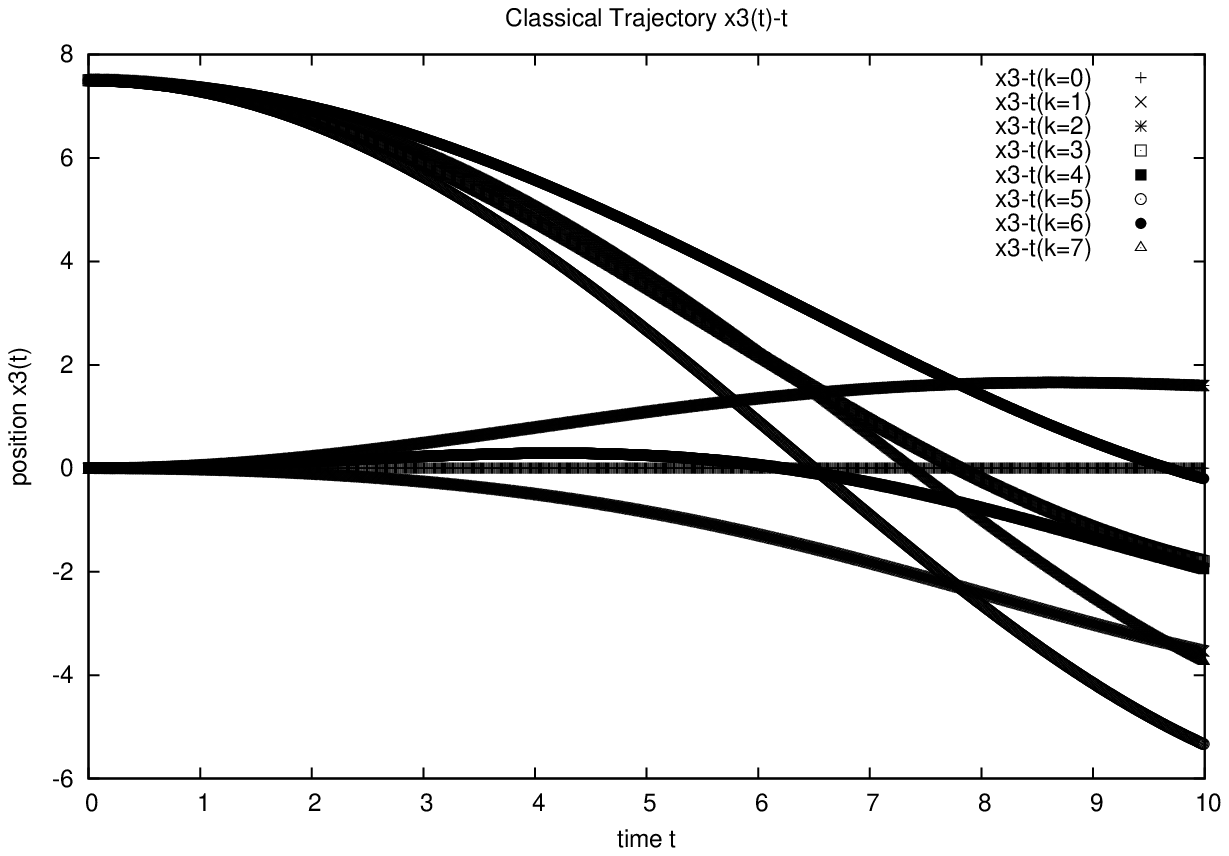}
\end{minipage}

\end{center}

\caption{ Time evolution of the reduced dencity of particle-3. (a) Time  t=4.005, when interference is still strong. (b) Time t=5.505, when the interference starts to damp. (c) Time t=10.005, when decoherence arises. (d) Classical trajectories of particle-3, position \(x_3\)(t) vs time t. Trajectories cross after time t=5.505.}
\label{f3}
\end{figure}

 Time evolutions of reduced density functions $\tilde{\rho}_1$,$\tilde{\rho}_2$,$\tilde{\rho}_3$ which are derived from quantum mechanical calculations are shown at upper three graphs in (Fig.\ref{f1}.- Fig.\ref{f3}.). As you can see, there are 2 packets in each graph. The one packet was at the origin ($x_1=0$ for Fig.\ref{f1}, etc.) initially. The another packet was at a distance ($x_1=d_1$ for Fig.\ref{f1}, etc.) initially. And there are 2 wave-like lines in each graph. The lower wave-like line is the quantum interference between these 2 packets. Disappearance of the interference means an emergence of classicality, that is decoherence. The upper wave-like line is the reduced density function for each particle. On the other hand, corresponding classical trajectories \(x_1(t),x_2(t),x_3(t)\) are shown at the bottom in each figure. 

 Lagrangian is Eq.(\( \ref{e1}  \)). Masses of three particles are the same, \(m\)=1.0. And \( \omega_{12}=0.305 \), \( \omega_{13}=0.1 \), \( \omega_{23}=0.202 \). Initial wave function for particle-1 is Eq.(\( \ref{e3}  \)) in which \(d_1\)=-5.0. For particle-2 and particle-3, \(d_2\)=6.0, \(d_3\)=7.5 respectively. And normalization constants \(\tilde{ N_1 },\tilde{ N_2 },\tilde{ N_3 }\) can be set 1, because we can use a numerical normalization. And \( \hbar=1.0\).

\begin{figure}[!]

\begin{center}

\begin{minipage}{.45\linewidth}
(a) t=2.005\\
\includegraphics[width=\linewidth]{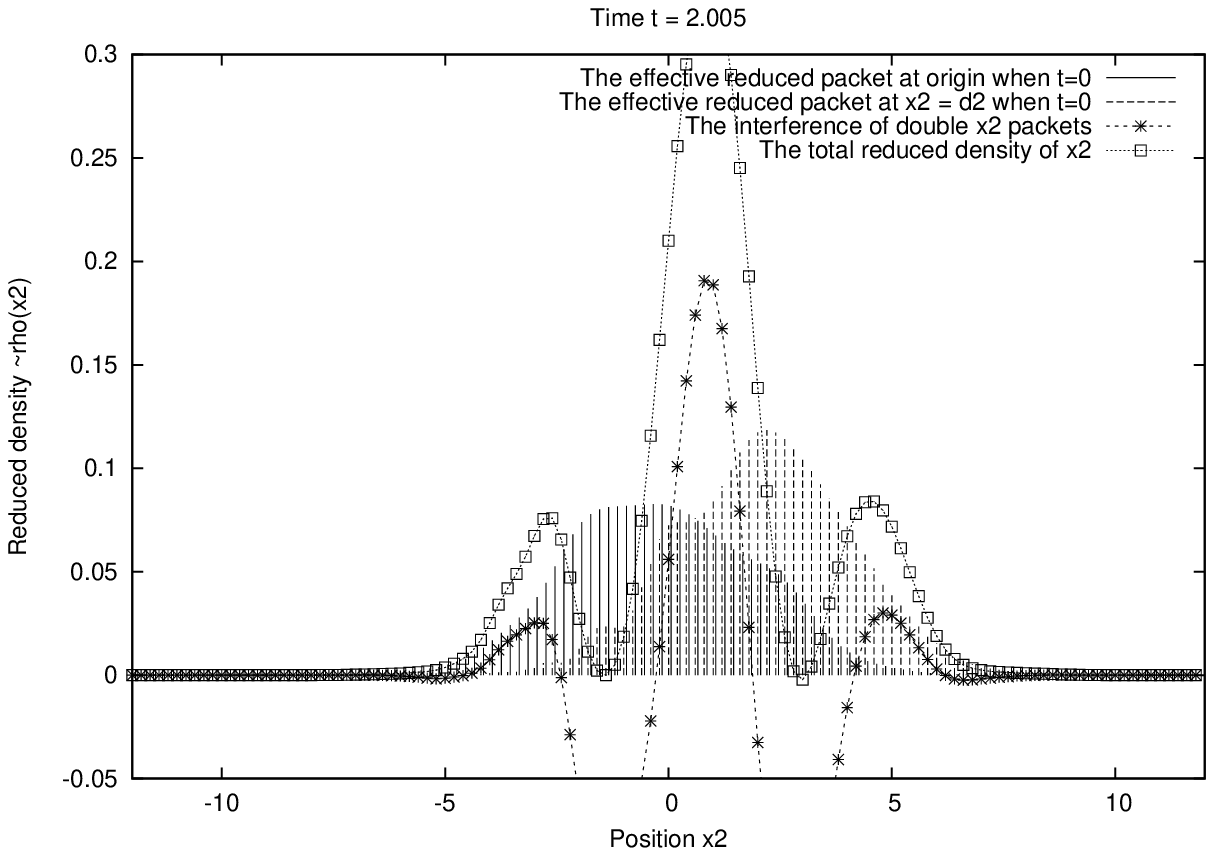}
\end{minipage}

\begin{minipage}{.45\linewidth}
(b) t=3.505\\
\includegraphics[width=\linewidth]{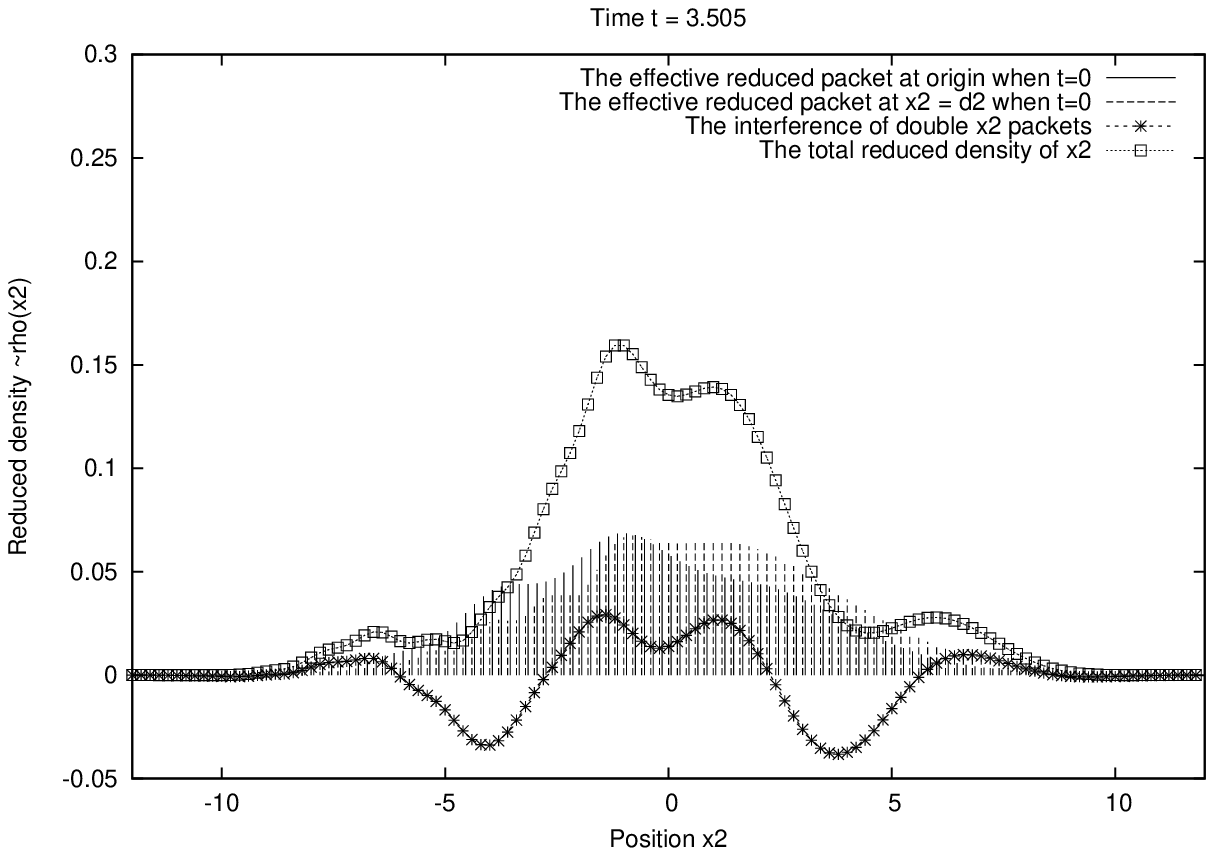}
\end{minipage}

\begin{minipage}{.45\linewidth}
(c) t=5.005\\
\includegraphics[width=\linewidth]{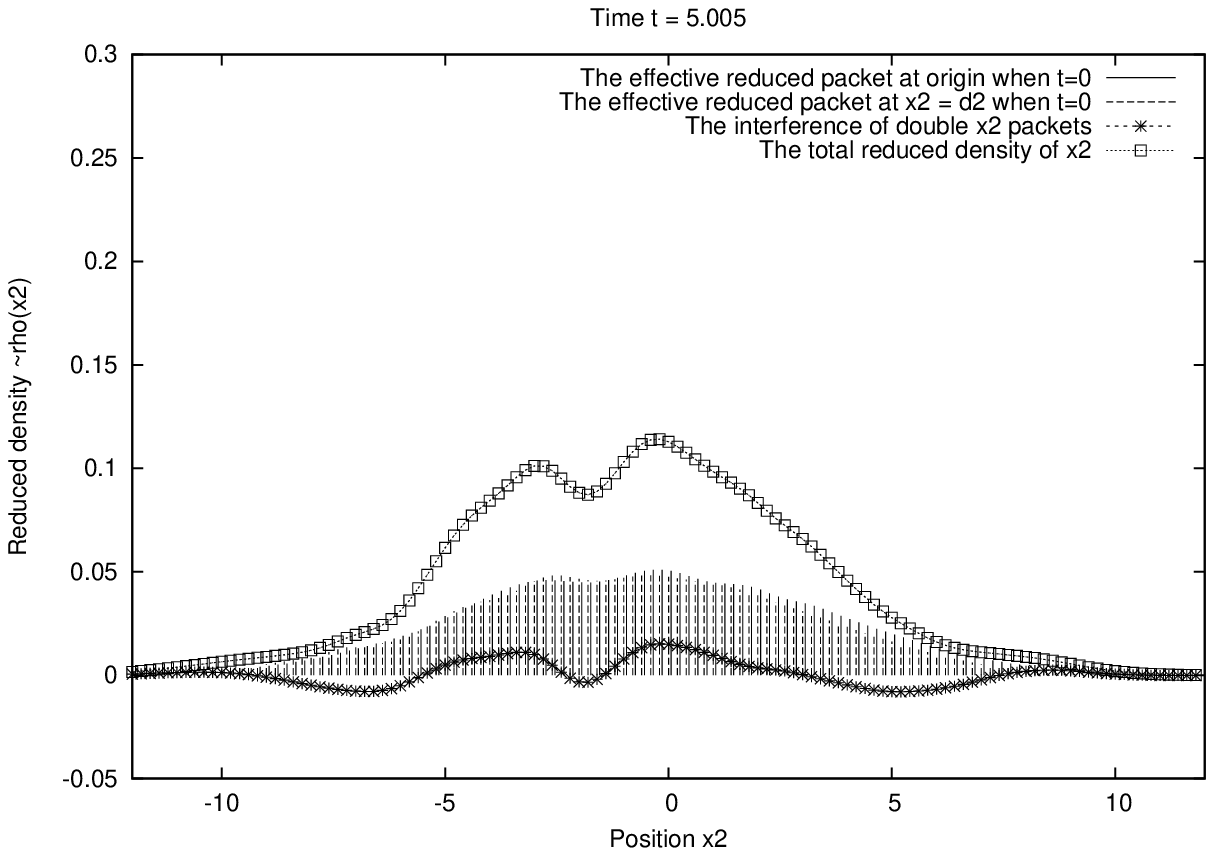}
\end{minipage}

\begin{minipage}{.5\linewidth}
(d) Classical trajectories \(x_2\)(t)\\
\includegraphics[width=\linewidth]{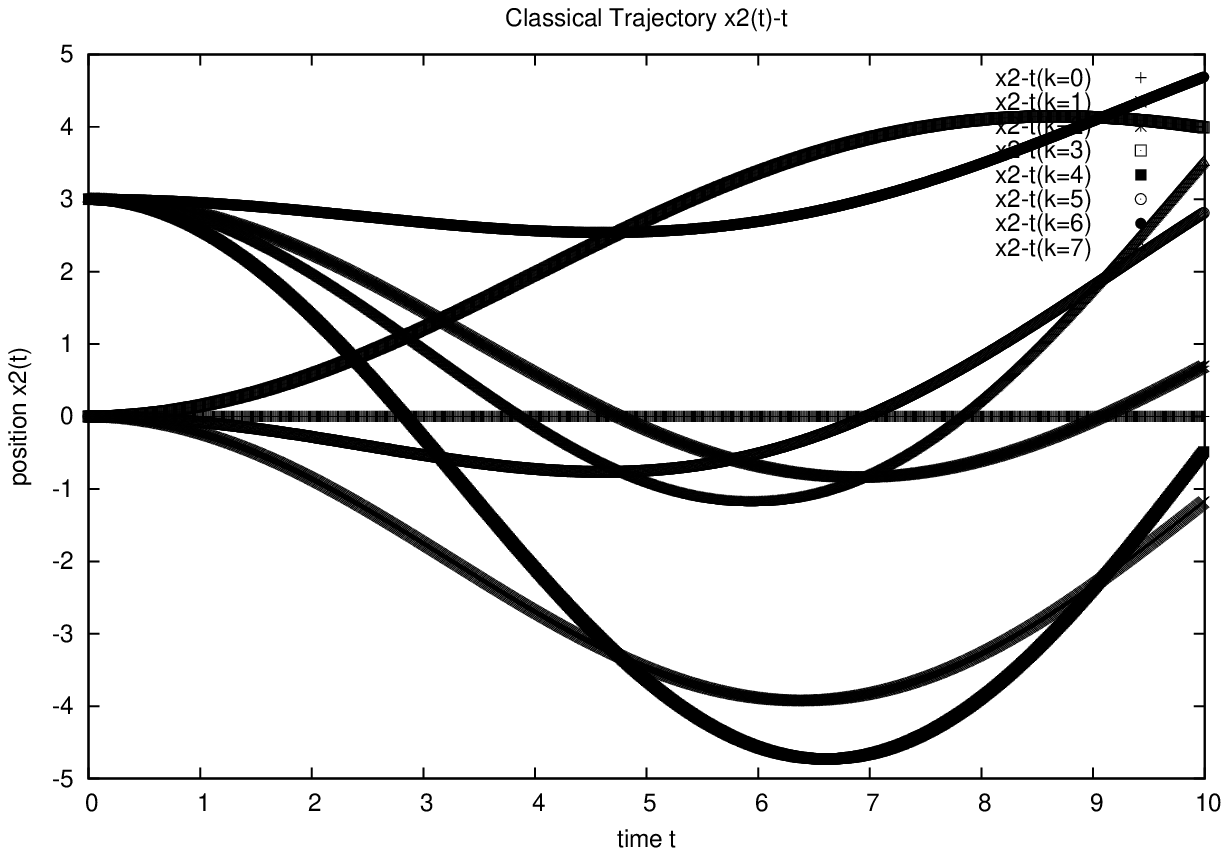}
\end{minipage}

\end{center}

\caption{ Time evolution of the reduced dencity of particle-2. (a) Time t=2.005, when interference is still strong. (b) Time t=3.505, when the interference starts to damp. (c) Time t=5.005, when decoherence arises. (d) Classical trajectories of particle-2, position \(x_2\)(t) vs time t. Trajectories cross after time t=2.005.}
\label{f2}
\end{figure}

\section{DISCUSSION}

 Above-mentioned result is rewritten as follows using state vectors schematically. The centers of packets above correspond to positions of particles in state vectors. \( |0 \ 0 \ 0 >\) is a state vector which expresses  "particle-1 is in origin, particle-2 is in origin, and particle-3 is in origin",  at a initial time \(t_0\). \quad \( |d_1 \ 0 \ 0 >\) is a state vector which expresses  "particle-1 is distant from origin, particle-2 is in origin, and particle-3 is in origin",  at \(t_0\). \( \quad \cdots \quad \)  \( |d_1 \ d_2 \ d_3 >\) is a state vector which expresses  "particle-1 is distant from origin, particle-2 is distant from origin, and particle-3 is distant from origin", at \(t_0\). In order to simplify this talk, the character in \(| \cdots >\) expresses only initial locations of three particles. Actually, the position of particles changes with time. This way of writing is for only distinguishing state vectors by the state of the initial time. 
 Three particles start to interact each other at initial time \(t_0\). At that time, each of three particles has two possible initial positions(the origin 0 and the distant position \(d_i (i=1,2,3)\)). Therefore a state vector of total three body system \( |\Phi >_{\mbox{total}} \) is a summation of \(2^3 = 8\) vectors as follows.
{\small  
\begin{eqnarray}
 |\Phi >_{\mbox{total}} &=&  |0 \ 0 \ 0> + |d_1 \ 0 \ 0> \nonumber \\
      &+& |0 \ d_2 \ 0> + |0 \ 0 \ d_3> + |d_1 \ d_2 \ 0> \nonumber \\
          &+& |0 \ d_2 \ d_3> + |d_1 \ 0 \ d_3> + |d_1 \ d_2 \ d_3> \nonumber \\ \label{e_dis1}
\end{eqnarray}
}
 By this research, it is suggested that wavefunction collapse by decoherence is not instant, and occurs gradually. With a time evolution, the state of the particle-2 is first decided by decoherence at the time t=2.0-5.0 (Fig.\ref{f2}).

 Here I have to emphasize some fact as follows. Although decoherence destroys interference between two quantum states of particle-2, it is considered to be powerless for the final determination of which state to come true for particle-2. Although the mechanism that finally decided which state will be realized is not known, anyway, only one state will be selected from either of two states. This time, the selected state for particle-2 is written as ''\(S_2\)''. (\(S_2\) is either \(d_2\) or \(0\).) 
If we assume that the unselected states will be vanished, then the state of total system becomes Eq.(\ref{e_dis3}).
{\small
\begin{equation}
  |\Phi >_{\mbox{total}} =  |0 \ S_2 \ 0> + |d_1 \ S_2 \ 0> + |0 \ S_2 \ d_3> + |d_1 \ S_2 \ d_3> \label{e_dis3}
\end{equation}
}
 Next, the state of the particle-1 is decided by decoherence at the time t=3.0-6.0 (Fig.\ref{f1}). Here, the selected state for particle-1 is written as ''\(S_1\)''. (\(S_1\) is either \(d_1\) or \(0\).)Then the state of total system becomes Eq.(\ref{e_dis4}).
{\small
\begin{equation}
  |\Phi >_{\mbox{total}} =  |S_1 \ S_2 \ 0> + |S_1 \ S_2 \ d_3> \label{e_dis4}
\end{equation}
}
 Finally, the state of the particle-3 is decided by decoherence at the time t=4.0-10.0 (Fig.\ref{f3}). Here, the selected state for particle-3 is written as ''\(S_3\)''. (\(S_3\) is either \(d_3\) or \(0\).) Then the state of total system becomes 
{\small
\begin{equation}
 |\Phi >_{\mbox{total}} = |S_1 \ S_2 \ S_3>. \label{e_dis2}
\end{equation}
}
This result implies that a closed system which has some kind of property (Probably, it is a dephasing by the intersection of classical trajectories$^{4}$.) gets classicality spontaneously without external environment or observer.

Therefore we can insist as follows. Our universe is a closed quantum mechanical system in principle and it does not have any external environments or observers. But it should have some kinetic property like this three Schr\"odinger cats model, therefore wavefunction of our universe collapses spontaneously and our universe gets classicality. The macroscopic quantum multiverse is difficult to exist, if this result is correct.

If you would like to confirm this result experimentally, an experiment is suggested. That you have to do is to throw this model and let it pass a double slit. Beautiful quantum interference stripe like the one arises when someone uses light, an electron, fullerene, will not arise. Since interference between two wave packets disappears before this model crashes into a screen.

\section{Conclusion}
 When decoherence arises with internal degrees of freedom only, each of degrees of freedom does not need to get classicality simultaneously. That is, determinations of their respective states occur gradually. In this three body model, after  respective decoherences for each of three particles have occurred, a state of a whole system of three particles seems to get classicality, too. That is, it is suggested that wavefunction of some kind of closed system (Probably, which includes a dephasing by the intersection of classical trajectories.) collapses spontaneously without any external environments. Therefore, it can be reasoned that the wavefunction  of the universe in which we live is also already collapsed by the same mechanism.

\section*{Acknowledgment}
 I would like to thank Prof.Fumihiko Sakata. I would like to thank my family, Yuzuki, Miyu and Yuko Ishikawa. 


%

\appendix
\section{Three Schr\"odinger cats model.}

  Here, We will use a simple model to discuss a possibility of quantum
 decoherence in a finite system. That is three bosonic particles tied up
 each other with three springs which have different frequencies. For
 this model, we will apply the Caldeira \& Leggett's technique. Our
 three particle model is a extreme reduction of their ``harmonic
 oscillator plus reservoir'' model.

 In this section, a system which includes 1-dimensional three particles tied with different springs will be transformed  to another coordinates system containing two uncoupled harmonic oscillators and a free particle. And a Feynman propagator for the new coordinates will be derived and be retranslated into the original three body system. Then we will get a propagator for harmonically bound three particles. 

 Next, respective initial wave functions for each particle will be prepared as a pair of Gaussian wave packets (Schr\"odinger cat state). The initial state of total three body system is the product of those Schr\"odinger cat states. Then it will start to turn into a entangled state of three particles by the propagator.

 In a isolated system, it is said that it's difficult to suppose that the interference term of wave function of isolated total system vanishes. But we will pay attention to the particle-1 only and integrate out other two degrees of freedom about other two particles. Then we will get a reduced density function for particle-1. This procedure corresponds to ignoring fine informations about other two particles and taking an average. Then, we will observe changes of the Gaussian wave packets of particle-1 and the interference term between them. \\

  \subsection{Classical model.}
 Classically, this model is solvable, that is a integrable system. Its Lagrangian is as follows.
\begin{equation}
   L  = \frac{m}{2} \dot{x_1}^2 +\frac{m}{2} \dot{x_2}^2 + \frac{m}{2}
   \dot{x_3}^2 -\frac{m}{2} \omega_{12}^2 (x_1-x_2)^2 -\frac{m}{2}
   \omega_{13}^2 (x_1-x_3)^2 -\frac{m}{2} \omega_{23}^2 (x_2-x_3)^2 \label{eqn007}
\end{equation}
  Applying the Euler-Lagrange equation
\begin{equation}
 \frac{\partial L}{\partial x_i} -\frac{d}{dt} \frac{\partial
  L }{\partial \dot{x_i}} = 0
\end{equation}
 to eq.(\(\ref{eqn007}\)), we can get three equations of motion.
\begin{equation}
 \frac{d^2}{dt^2}
     \left( \begin{array}{c}
         x_1 \\ x_2 \\ x_3        
            \end{array} \right)
 = \left( \begin{array}{ccc}
      -(\omega_{12}^2 + \omega_{13}^2) & \omega_{12}^2 & \omega_{13}^2 \\
      \omega_{12}^2 & -(\omega_{12}^2 + \omega_{23}^2) & \omega_{23}^2 \\
      \omega_{13}^2 & \omega_{23}^2 & -(\omega_{13}^2 + \omega_{23}^2)
          \end{array} \right)
     \left( \begin{array}{c}
         x_1 \\ x_2 \\ x_3        
            \end{array} \right)
\end{equation}
 Rewriting this,
\begin{equation}
  \frac{d^2}{dt^2} \vecX_{(t)} = \vecW \vecX_{(t)}
\end{equation}
 and using a time-independent matrix \( \vecP \) then,
\begin{equation}
  \frac{d^2}{dt^2} ( \vecP \vecX_{(t)} ) = \vecP \vecW \vecP^{-1} (
   \vecP \vecX_{(t)} ) \label{eqn010}
\end{equation}
 Here, we can select \( \vecP \) for making \( \vecP \vecW \vecP^{-1} \)
 diagonal, thus we can get three uncoupled differential equations. I will
 show this. Eigen values of \( \vecW \), \( \lambda \) satisfy an equation
\begin{equation}
 -\left\{ \ \lambda ^3 + 2(\omega_{12}^2 + \omega_{13}^2 + \omega_{23}^2)
   \lambda^2 + 3(\omega_{12}^2 \omega_{13}^2 +\omega_{13}^2
   \omega_{23}^2+\omega_{12}^2 \omega_{23}^2) \lambda \ \right\} \equiv
 0 \label{eqn012}
\end{equation}
. We can derive its solutions \( \lambda_1, \lambda_2, \lambda_3 \), and
obviously 0 is a solution, so we set \( \lambda_3 \) to be 0. And we define
\begin{equation}
  \vecP \vecW \vecP^{-1} = \left( 
     \begin{array}{ccc}     
        \lambda_1 & \ & \ \\
        \ & \lambda_2 & \ \\
        \ & \ & \lambda_3
     \end{array}  \right)
    = \left(
     \begin{array}{ccc}     
        \lambda_1 & 0 & 0 \\
        0 & \lambda_2 & 0 \\
        0 & 0 & 0
     \end{array}  \right)
    \equiv \vecLambda 
\end{equation}
\begin{equation}
  \vecP \vecX_{(t)} \equiv \vecZ_{(t)}
\end{equation}
. Then, equation (\ref{eqn010}) becomes to 
\begin{equation}
  \frac{d^2}{dt^2} \vecZ _{(t)} = \vecLambda \vecZ _{(t)}
\end{equation}
, that is, we can get three independent differential equations as follows.
\begin{equation}
  \frac{d^2}{ d t^2} \left( 
 \begin{array}{c} z_1 \\ z_2 \\ z_3  \end{array}
  \right) = \left( 
 \begin{array}{ccc}
   \lambda_1 & \ & \ \\ \ & \lambda_2 & \ \\ \ & \ & 0
 \end{array}
  \right) \left(
 \begin{array}{c} z_1 \\ z_2 \\ z_3  \end{array}
  \right) = \left(
 \begin{array}{c} \lambda_1 z_1 \\ \lambda_2 z_2 \\ 0  \end{array}
  \right) \label{eqn017}
\end{equation}
 Here we define 
\begin{eqnarray}
  {\mit \Delta} \omega^2 &\equiv& \sqrt{ \omega_{12}^4 -\omega_{12}^2
   \omega_{13}^2 +\omega_{13}^4 -\omega_{13}^2 \omega_{23}^2 +\omega_{23}^4
   -\omega_{23}^2 \omega_{12}^2 } \nonumber \\
  &=& \sqrt{ \frac{1}{2} \left\{ (\omega_{12}^2-\omega_{13}^2)^2
   +(\omega_{13}^2-\omega_{23}^2)^2 +(\omega_{23}^2-\omega_{12}^2)^2
		       \right\} } \nonumber \\
  &=& \sqrt{(\omega_{12}^2 +\omega_{13}^2 +\omega_{23}^2 )^2
   -3\omega_{12}^2 \omega_{13}^2 -3\omega_{13}^2 \omega_{23}^2 -3\omega_{23}^2 \omega_{12}^2}
\end{eqnarray}
 and from equation (\ref{eqn012}), \(\lambda_1,\lambda_2\) are 
\begin{equation}
  \left\{ \begin{array}{l}
    \lambda_1 = -\omega_{12}^2 -\omega_{13}^2 -\omega_{23}^2 +{\mit \Delta}
     \omega^2 < 0 \\
    \lambda_2 = -\omega_{12}^2 -\omega_{13}^2 -\omega_{23}^2 -{\mit \Delta}
     \omega^2 < 0
   \end{array} \right.
\end{equation}
. Paying attention to their signs, we can solve the equation (\ref{eqn017}).
Then we get 
\begin{eqnarray}
  && z_{1(t)} = A_1 \sin{\Omega_1 t} +B_1 \cos{\Omega_1 t} \quad \nonumber \\
  && z_{2(t)} = A_2 \sin{\Omega_2 t} +B_2 \cos{\Omega_2 t} \quad \nonumber \\
  && z_{3(t)} = C_1 t +C_2 \quad \label{eqn20140330_1}
\end{eqnarray}
, where
\begin{equation}
  \Omega_1 \equiv \sqrt{-\lambda_1}, \  \Omega_2 \equiv \sqrt{-\lambda_2}
\end{equation}
 and \( A_1,A_2,B_1,B_2,C_1,C_2 \) are integral constants. They depend on
 initial conditions of \( \vecX (t) \) and \( \dot{\vecX}(t) \). 
For example, from eq.(\ref{eqn20140330_1}),
\begin{equation}
 A_1 =  \frac{\dot{z}_{1(t_0)}}{\Omega_1}, A_2 =  \frac{\dot{z}_{2(t_0)}}{\Omega_2}, B_1 = z_{1(t_0)},  B_2 = z_{2(t_0)}, C_1 = \dot{z}_{3(t_0)}, C_2 = z_{3(t_0)} \label{eqn20140419_1}
\end{equation}
 when the initial time \(t_0 = 0\). And \(z_{i(0)},\dot{z}_{i(0)} (i=1,2,3)\) will be written with  \(x_{i(0)},\dot{x}_{i(0)} (i=1,2,3)\), when you use eq.(\ref{eqn028}) later.

The classical trajectory \( \vecX (t) \) is 
\begin{equation}
 \vecX _{(t)} = \vecP^{-1} \vecZ_{(t)}
\end{equation}
. We define \( \vecW \)'s three eigen vectors \( { \bf p_1,p_2,p_3 } \), they are corresponding to their respective eigen value \( \lambda_1,\lambda_2,0 \).
\begin{equation}
  \vecP^{-1} = \left( { \bf p_1 \ p_2 \ p_3 } \right) = 
 \left( \begin{array}{ccc}
  \xi_1 & \xi_2 & 1 \\ \eta_1 & \eta_2 & 1 \\ \zeta_1 & \zeta_2 & 1
 \end{array} \right)
= 
 \left( \begin{array}{ccc}
  \xi_1 & \xi_2 & 1 \ \\ \eta_1 & \eta_2 & 1 \ \\ -\xi_1-\eta_1 &
   -\xi_2-\eta_2 & 1 \
 \end{array} \right)
\end{equation}
 Here we difined
\begin{equation}
 \begin{array}{ll}
  \xi_1 = \omega_{12}^2 \omega_{23}^2 - \omega_{13}^2 (\omega_{13}^2
   -{\mit \Delta} \omega^2) \quad & \xi_2 = \omega_{12}^2 \omega_{23}^2 - \omega_{13}^2 (\omega_{13}^2
   +{\mit \Delta} \omega^2) \\
  \eta_1 = \omega_{12}^2 \omega_{13}^2 - \omega_{23}^2 (\omega_{23}^2
   -{\mit \Delta} \omega^2) \quad & \eta_2 = \omega_{12}^2 \omega_{13}^2 - \omega_{23}^2 (\omega_{23}^2
   +{\mit \Delta} \omega^2) \\
  \zeta_1 = -\xi_1 -\eta_1 \quad & \zeta_2 = -\xi_2 -\eta_2
 \end{array}
\end{equation}
 Using the formulae above, we can get a classical solution of \( \vecX
 \)(t) finally.
\begin{equation}
 \vecX_{(t)} = 
  \left( \begin{array}{c} 
   x_{1(t)} \\ x_{2(t)} \\ x_{3(t)}
  \end{array} \right) =
  \left( \begin{array}{c} 
    \xi_1 z_{1(t)} +\xi_2 z_{2(t)} +z_{3(t)} \\
    \eta_1 z_{1(t)} +\eta_2 z_{2(t)} +z_{3(t)} \\
    -(\xi_1+\eta_1) z_{1(t)} -(\xi_2+\eta_2) z_{2(t)} +z_{3(t)} \\
  \end{array} \right)  \label{eqn025}
\end{equation}
 And with new constant \( \Delta \equiv \eta_2 \xi_1 -\eta_1 \xi_2 \), you can write \( \vecP \).
\begin{equation}
  \vecP = \frac{1}{3\Delta} 
   \left( \begin{array}{ccc}
    2\eta_2 +\xi_2 & -\eta_2 -2\xi_2 & -\eta_2 +\xi_2 \\
    -2\eta_1 -\xi_1 & \eta_1 +2\xi_1 & \eta_1 -\xi_1 \\
    \Delta & \Delta & \Delta
   \end{array} \right) \label{eqnPPP}
\end{equation}
  Then from
\begin{equation}
  \vecZ_{(t)} = \vecP \vecX_{(t)}
\end{equation}
 , we can get a formula for transformation to normal coordinates \(
 \vecZ (t) \).
\begin{equation}
  \vecZ_{(t)} = 
  \left( \begin{array}{c} 
   z_{1(t)} \\ z_{2(t)} \\ z_{3(t)}
  \end{array} \right) = \frac{1}{3\Delta}
  \left( \begin{array}{c} 
    (2\eta_2 +\xi_2) x_{1(t)} +(-\eta_2 -2\xi_2) x_{2(t)} +(-\eta_2 +\xi_2) x_{3(t)} \\
    (-2\eta_1 -\xi_1) x_{1(t)} +(\eta_1 +2\xi_1) x_{2(t)} +(\eta_1 -\xi_1)x_{3(t)} \\
    \Delta x_{1(t)} +\Delta x_{2(t)} +\Delta x_{3(t)} \\
  \end{array} \right) \label{eqn028}
\end{equation}
 These formulae are very useful for evaluation of path integrals later.

 \subsection{Derivation of a propagator}

  In this section, a Feynman propagator for this model is derived. It describes
 an time evolution of wave functions. It is difficult to
 derive a propagator in the original coordinates \( \vecX (t) \), so 
 Lagrangian which contains \( \vecX (t) \)  is transformed into another Lagrangian which contains the normal coordinates \( \vecZ (t) \) . There are two uncoupled harmonic oscillators and a free particle in coordinates \( \vecZ (t) \).

 The transformation formulae are given in eq.(\ref{eqn025}).  The normal coordinates \( \vecZ (t) \) are substituted into the original cordinates \( \vecX (t) \) in the original Lagrangian eq.(\(\ref{eqn007}\)), then we get a equivalent Lanrangian in the normal coordinates \( \vecZ (t) \).

\begin{equation}
  L = \frac{m_1}{2} \dot{z}^2_{1(t)} +\frac{m_2}{2} \dot{z}^2_{2(t)}
   +\frac{m_3}{2} \dot{z}^2_{3(t)} -\frac{m_1}{2} \omega_1^2 z^2_{1(t)}
   -\frac{m_2}{2} \omega_2^2 z^2_{2(t)}
\end{equation}
 where 
\begin{equation}
  m_1 \equiv 2m(\xi_1^2 +\xi_1\eta_1 +\eta_1^2), 
  \quad m_2 \equiv 2m(\xi_2^2 +\xi_2\eta_2 +\eta_2^2),
  \quad m_3 \equiv 3m
\end{equation}
\begin{equation} \left\{
 \begin{array}{c}
  \omega_1^2 \equiv  \frac{m}{m_1} \{ \
   w_1^2 (2\xi_1^2 -\xi_1\eta_1 -\eta_1^2)
   +w_2^2(-\xi_1^2 -\xi_1\eta_1 +2\eta_1^2)
   +w_3^2(2\xi_1^2 +5\xi_1\eta_1 +2\eta_1^2) \ \} \\
  \omega_2^2 \equiv  \frac{m}{m_2} \{ \
   w_1^2 (2\xi_2^2 -\xi_2\eta_2 -\eta_2^2)
   +w_2^2(-\xi_2^2 -\xi_2\eta_2 +2\eta_2^2)
   +w_3^2(2\xi_2^2 +5\xi_2\eta_2 +2\eta_2^2) \ \} \\
  ( \ w_1^2 = \omega_{12}^2+\omega_{13}^2, \quad  w_2^2 =
   \omega_{12}^2+\omega_{23}^2, \quad  w_3^2 =
   \omega_{13}^2+\omega_{23}^2 \ )
  \end{array} \right.
\end{equation}
  So, we can decouple it with each variables. 
\begin{eqnarray}
  &&L_1(z_{1(t)},t) \equiv \frac{m_1}{2} \dot{z}^2_{1(t)} -\frac{m_1}{2} \omega_1^2
   z^2_{1(t)}, \
  L_2(z_{2(t)},t) \equiv \frac{m_2}{2} \dot{z}^2_{2(t)} -\frac{m_2}{2} \omega_2^2
  z^2_{2(t)}, \ \nonumber \\
  && \quad L_3(z_{3(t)},t) \equiv \frac{m_3}{2} \dot{z}^2_{3(t)}  \quad
   : L = L_1 +L_2 + L_3 
\end{eqnarray}
 For these Lagrangians, we can get classical action integrals which are summed up from an initial time \( t_0 \) to an arbitrary time t.
\begin{eqnarray}
 &\mbox{ }& S^{(cl)}(\vecZ_{(t)},t:\vecZ_{(t_0)},t_0) 
  = \int_{t_0}^t L_{1(\tau)} d\tau +\int_{t_0}^t L_{2(\tau)} d\tau
  +\int_{t_0}^t L_{3(\tau)} d\tau \nonumber \\
 &\mbox{ }& \quad\quad \equiv S^{(cl)}_1(z_{1(t)},t:z_{1(t_0)},t_0)
  +S^{(cl)}_2(z_{2(t)},t:z_{2(t_0)},t_0)
  +S^{(cl)}_3(z_{3(t)},t:z_{3(t_0)},t_0) \nonumber \\
 && \
\end{eqnarray}
 Here respective actions are as follows.
\begin{eqnarray}
  && S^{(cl)}_1 = \frac{m_1 \omega_1}{2\sin{[\omega_1(t-t_0)]}} 
       \left\{  \cos{[\omega_1(t-t_0)]} ( z^2_{1(t)} +z^2_{1(t_0)} )
	-2z_{1(t)}z_{1(t_0)}  \right\} \\
  && S^{(cl)}_2 = \frac{m_2 \omega_2}{2\sin{[\omega_2(t-t_0)]}} 
       \left\{  \cos{[\omega_2(t-t_0)]} ( z^2_{2(t)} +z^2_{2(t_0)} )
	-2z_{2(t)}z_{2(t_0)}  \right\} \\
  && S^{(cl)}_3 = \frac{m_3 ( z_{3(t)} -z_{3(t_0)} )^2}{2(t-t_0)}
\end{eqnarray}
 Using these formulae, we can get the propagator
 in the system \( \vecZ(t) \). In following, \( (t) \) is often omitted, \( \vecZ =\vecZ(t) \),\( \vecX =\vecX(t) \). And \( \vecZ_0 =\vecZ(t_0) \),\( \vecX_0 =\vecX(t_0) \).
\begin{eqnarray}
  U(\vecZ,t:\vecZ_0,t_0) &=& \int_{\vecZ(t_0)=\vecZ_0}^{\vecZ(t)=\vecZ}
   D\vecZ(\tau) \exp \left\{ \frac{i}{\hbar}
		      S(\vecZ,t:\vecZ(\tau),\tau:\vecZ_0,t_0)   \right\}   \\
  &\propto& \exp \left\{\frac{i}{\hbar} S^{(cl)}(\vecZ,t:\vecZ_0,t_0)
		 \right\} \label{eqn038}
\end{eqnarray}
 Here \( D\vecZ \equiv Dz_1 Dz_2 Dz_3 \) means path integrals for three
 variables in system \( \vecZ \). Though the action integral \(
 S(\vecZ,t:\vecZ(\tau),\tau:\vecZ_0,t_0) \) depends on these integral
 paths and does not always follow the principle of minimum action, for
 free particles and for harmonic oscillators, it is known that the
 result of path integrals is in proportion to the value of saddle point
 of their integrands, that is, it comes to equation (\ref{eqn038}). Maybe its
 proportional factor depends on the initial time \(t_0\) and the final time
 t, but here I omit it. Because it is for a simplicity, and we can use computational normalization after.

 Now we assume that the propagator in \( \vecZ \) is equivalent to the
 propagator in original system \( \vecX \). Transforming it with equation
 (\ref{eqn028}), then we get equation (\ref{eqn039}).

\begin{eqnarray}
 &\mbox{ }& U(\vecX,t:\vecX_0,t_0) = U(\vecZ,t:\vecZ_0,t_0) \nonumber \\
  &\mbox{ }& \qquad \propto \exp \biggl[ \frac{i}{\hbar} \left\{ A_1 x_{1(0)}^2 +A_2 x_{2(0)}^2 
  +A_3 x_{3(0)}^2 +B_{12} x_{1(0)} x_{2(0)} \right. \biggr.\nonumber \\
   &\mbox{ }& \qquad\qquad \biggl. \left. +B_{23} x_{2(0)} x_{3(0)} +B_{13} x_{1(0)} x_{3(0)} +C_1 x_{1(0)}
  +C_2 x_{2(0)} +C_3 x_{3(0)} +D \ \right\} \biggr] \label{eqn039}
   \nonumber \\
  && \
\end{eqnarray}
  New variables in equation
 (\ref{eqn039}) are defined as follows. They are different from coefficients \( A_1 - C_2 \) in eq.(\ref{eqn20140330_1}) and (\ref{eqn20140419_1}).  Matrix elements of the transformation matrix \( \vecP \) in equation (\ref{eqnPPP}) are rewritten as
\begin{equation}
 \begin{array}{lll}
   a_1 = \frac{1}{3\Delta}(2\eta_2+\xi_2) & 
   a_2 = \frac{1}{3\Delta}(-\eta_2-2\xi_2) & 
   a_3 = \frac{1}{3\Delta}(-\eta_2+\xi_2) \\
   b_1 = \frac{1}{3\Delta}(-2\eta_1-\xi_1) & 
   b_2 = \frac{1}{3\Delta}(\eta_1+2\xi_1) & 
   b_3 = \frac{1}{3\Delta}(\eta_1-\xi_1) \\
   c_1 = \frac{1}{3} & c_2 = \frac{1}{3} & c_3 = \frac{1}{3}
 \end{array}
\end{equation}
 , where constant (\( \Delta = \eta_2\xi_1-\eta_1\xi_2 \)). Then we can get real coefficients \( A_1
 - D \) in equation (\ref{eqn039}) as follows. 

\begin{eqnarray}
  &\mbox{ }& A_i = \frac{m_1\omega_1}{2}\cot[\omega_1(t-t_0)]a_i^2
   +\frac{m_2\omega_2}{2}\cot[\omega_2(t-t_0)]b_i^2
   +\frac{m_3}{2(t-t_0)} c_i^2 \nonumber \\ 
  &\mbox{ }& B_{ij} = m_1\omega_1 \cot [\omega_1 (t-t_0)]a_i a_j +m_2\omega_2 \cot
   [\omega_2 (t-t_0)]b_i b_j +\frac{m_3}{(t-t_0)} c_i c_j   \nonumber \\
  &\mbox{ }& C_i(\vecX) = -\frac{m_1 \omega_1}{\sin[\omega_1(t-t_0)]}
     \left( a_1 x_1 + a_2 x_2 + a_3 x_3 \right) a_i \nonumber \\
  &\mbox{ }& \qquad\qquad\qquad\qquad\qquad -\frac{m_2 \omega_2}{\sin[\omega_2(t-t_0)]}
     \left( b_1 x_1 + b_2 x_2 + b_3 x_3 \right) b_i \nonumber \\
  &\mbox{ }& \qquad\qquad\qquad\qquad\qquad\qquad\qquad\qquad -\frac{m_3}{(t-t_0)}
     \left( c_1 x_1 + c_2 x_2 + c_3 x_3 \right) c_i \nonumber \\
  &\mbox{ }& D(\vecX) = \frac{m_1 \omega_1}{2} \cot[\omega_1(t-t_0)](a_1 x_1 +
  a_2 x_2 + a_3 x_3)^2  \nonumber \\
  &\mbox{ }& \qquad\qquad\qquad\qquad\qquad +\frac{m_2 \omega_2}{2} \cot[\omega_2(t-t_0)](b_1 x_1 + b_2 x_2 + b_3
   x_3)^2  \nonumber \\
  &\mbox{ }& \qquad\qquad\qquad\qquad\qquad\qquad\qquad\qquad\quad +\frac{m_3}{2(t-t_0)}(c_1 x_1 + c_2 x_2 + c_3 x_3)^2 \nonumber \\
  &\mbox{ }&
\end{eqnarray}
 Here \( (i,j = 1,2,3 )\), and index \((t)\) is omitted, that is  \((x_{i(t)}=x_i)\).

 \subsection{Derivation of wave function and calculation
  of reduced density function.}
 Using the propagator above, we can write a evolution of a wave
  function of whole system as follows. 
\begin{equation}
  \psi(\vecX,t) = \int_{-\infty}^{\infty} d\vecX_0 \ U(\vecX,t:\vecX_0,t_0)
   \ \psi(\vecX_0,t_0) \label{eqn045}
\end{equation}
 The initial wave function \(\psi(\vecX_0,t_0)\) is a product of initial wave functions of each particle. 

\begin{equation}
 \psi(\vecX_0,t_0) = \psi_1(x_{1(0)},t_0) \ \psi_2(x_{2(0)},t_0) \
  \psi_3(x_{3(0)},t_0) \label{eqn046}
\end{equation}
 Here, each initial state is the Schr\"odinger cat state.
\begin{equation}
 \psi_1(x_{1(0)},t_0) = \tilde{ N_1 } \left[
				       \exp\left\{-\frac{x_{1(0)}^2}{4\sigma_1^2}\right\}+\exp\left\{-\frac{(x_{1(0)}-d_1)^2}{4\sigma_1^2}\right\}
				      \right]  \quad \mbox{etc...} \label{eqn047}
\end{equation}
 With (\(\ref{eqn039}\)),(\(\ref{eqn046}\)) and (\(\ref{eqn047}\)), equation
 (\(\ref{eqn045}\)) comes to equation (\(\ref{eqn048}\)).

\begin{eqnarray}
 && \qquad\ \psi(\vecX,t) \propto \int_{-\infty}^{\infty} dx_{1(0)}
  \int_{-\infty}^{\infty} dx_{2(0)} \int_{-\infty}^{\infty} dx_{3(0)}
  \nonumber \\
 &&  \times \exp \left[ \frac{i}{\hbar} \bigl\{  A_1 x_{1(0)}^2 +A_2 x_{2(0)}^2
			       +A_3 x_{3(0)}^2  
 +B_{12} x_{1(0)} x_{2(0)} +B_{23} x_{2(0)} x_{3(0)}  \bigr. \right. \nonumber \\ 
 && \qquad\qquad\qquad\qquad\qquad\qquad\qquad
  \biggl. \bigl. +B_{13} x_{1(0)} x_{3(0)} +C_1 x_{1(0)} +C_2 x_{2(0)} +C_3 x_{3(0)} +D  \bigr\}
  \biggr] \nonumber \\
 && \times \biggl[ \exp \left\{
			 -\frac{x_{1(0)}^2}{4\sigma_1^2}-\frac{x_{2(0)}^2}{4\sigma_2^2}-\frac{x_{3(0)}^2}{4\sigma_3^2}  \right\} +\exp \left\{ -\frac{(x_{1(0)}-d_1)^2}{4\sigma_1^2}-\frac{x_{2(0)}^2}{4\sigma_2^2}-\frac{x_{3(0)}^2}{4\sigma_3^2}  \right\} + \cdots \nonumber \\
 && \qquad\qquad \cdots +\exp \left\{
			       -\frac{(x_{1(0)}-d_1)^2}{4\sigma_1^2}-\frac{(x_{2(0)}-d_2)^2}{4\sigma_2^2}-\frac{(x_{3(0)}-d_3)^2}{4\sigma_3^2}  \right\} \biggr] \label{eqn048}
\end{eqnarray}
 The latter \([ \cdots ]\) of this formula means 8 Gaussian packets in
 (\( x_1,x_2,x_3 \)) space. Each packet changes by propagator. For
 evaluating their analytic forms, integration with three variables
 is needed. With new complex coefficients \(\breve{A}_1 - \breve{D}
 \), it turns into
\begin{eqnarray}
 && \qquad\ \psi(\vecX,t) = \sum_{k=0-7} \psi^{(k)} (\vecX,t) \\
 && \qquad\ \psi^{(k)}(\vecX,t) \propto \int_{-\infty}^{\infty} dx_{1(0)}
  \int_{-\infty}^{\infty} dx_{2(0)} \int_{-\infty}^{\infty} dx_{3(0)}
  \nonumber \\
 &&  \times \exp \left[   -\breve{A}_1 x_{1(0)}^2 -\breve{A}_2 x_{2(0)}^2
			       -\breve{A}_3 x_{3(0)}^2  
 +\breve{B}_{12} x_{1(0)} x_{2(0)} +\breve{B}_{23} x_{2(0)} x_{3(0)}  \right. \nonumber \\ 
 && \qquad\qquad\qquad\qquad\qquad \biggl. +\breve{B}_{13} x_{1(0)}
  x_{3(0)} +\breve{C}^{(k)}_1 x_{1(0)} +\breve{C}^{(k)}_2 x_{2(0)}
  +\breve{C}^{(k)}_3 x_{3(0)} +\breve{D}^{(k)}  \biggr] \nonumber \\
 && \
\end{eqnarray}
 (\( k=0-7 \)), which is the Gaussian integrals we have to solve .
\begin{eqnarray}
 && \breve{A}_1 = \frac{1}{4\sigma_1^2} - \frac{i}{\hbar}A_1, \quad 
  \breve{B}_{12} = \frac{i}{\hbar}B_{12}, \quad  \breve{C}^{(k)}_1(\vecX) =
  \frac{d^{(k)}_1}{2\sigma_1^2} +\frac{i}{\hbar}C_1(\vecX)
\quad \mbox{etc...} \nonumber \\
 && \qquad\qquad \breve{D}^{(k)}(\vecX) = -\frac{d^{(k)2}_1}{4\sigma_1^2}
  -\frac{d^{(k)2}_2}{4\sigma_2^2} -\frac{d^{(k)2}_3}{4\sigma_3^2} +\frac{i}{\hbar}D(\vecX)
\end{eqnarray}
 Here \(d^{(k)}_1, d^{(k)}_2,d^{(k)}_3 \) are defined for each (\(k=0-7\)) as follows.
\begin{equation} \left(
 \begin{array}{rrcccl}
   \quad & [& d^{(k)}_1 & d^{(k)}_2 & d^{(k)}_3 &] \\
   k = 0 & [& \ 0 & \ 0 & \ 0 &] \\
       1 & [& d_1 & \ 0 & \ 0 &] \\
       2 & [& \ 0 & d_2 & \ 0 &] \\
       3 & [& \ 0 & \ 0 & d_3 &] 
 \end{array}
 \quad
 \begin{array}{rrcccl}
       \ & \ & \quad \ & \quad \ & \quad \  & \ \\
       4 & [& d_1 & d_2 & \ 0 &] \\
       5 & [& d_1 & \ 0 & d_3 &] \\
       6 & [& \ 0 & d_2 & d_3 &] \\
       7 & [& d_1 & d_2 & d_3 &] 
 \end{array} \right) \label{eqn052}
\end{equation}
  Evaluating this Gaussian integrals, we see
\begin{equation}
  \psi^{(k)}(\vecX,t) \propto \sqrt{\frac{\pi^3}{\Delta}} \exp \left[
 \frac{1}{16\Delta}\Phi^{(k)}(\vecX,t) +\breve{D}^{(k)}(\vecX,t) \right] \label{eqn053}
\end{equation}
 , where new \(\Delta = \Delta(t)\) is defined as follows.
\begin{eqnarray}
  && \Delta(t) = \breve{A}_1 \breve{A}_2 \breve{A}_3
   -\frac{1}{4}( \breve{A}_2 \breve{B}^2_{13} +\breve{A}_3
   \breve{B}^2_{12} +\breve{A}_1 \breve{B}^2_{23} )
   -\frac{1}{4}\breve{B}_{12} \breve{B}_{13} \breve{B}_{23} \\[5pt]
  && \qquad \ \equiv \Re e \Delta(t) +i \ \Im m \Delta(t) \\[5pt]
  && \quad \Re e \Delta(t) = \frac{1}{4^3 \sigma_1^2 \sigma_2^2
   \sigma_3^2}-\frac{1}{4\hbar^2}\left( \frac{A_2 A_3}{\sigma_1^2}
  +\frac{A_3 A_1}{\sigma_2^2} +\frac{A_1 A_2}{\sigma_3^2} \right)
   \nonumber \\
  && \qquad\qquad\qquad\qquad  +\frac{1}{16\hbar^2} \left(
   \frac{B_{23}^2}{\sigma_1^2} +\frac{B_{13}^2}{\sigma_2^2}
   +\frac{B_{12}^2}{\sigma_3^2} \right) \\
  && \quad \Im m \Delta(t) = \frac{1}{\hbar^3}A_1 A_2 A_3
   -\frac{1}{16\hbar}\left( \frac{A_3}{\sigma_1^2 \sigma_2^2}
    +\frac{A_1}{\sigma_2^2 \sigma_3^2} +\frac{A_2}{\sigma_3^2
    \sigma_1^2} \right) \nonumber \\
  && \qquad\qquad\qquad\qquad -\frac{1}{4\hbar^3}(A_1 B_{23}^2 +A_2 B_{13}^2
    +A_3 B_{12}^2) +\frac{1}{4\hbar^3}B_{12}B_{13}B_{23} \nonumber \\
  &\mbox{}&
\end{eqnarray}
 , while
\begin{eqnarray}
  && \Phi^{(k)}(\vecX,t) = 4( \ \breve{A}_2 \breve{A}_3 \breve{C}^{2}_1(\vecX)
   +\breve{A}_1 \breve{A}_3 \breve{C}^{2}_2(\vecX) +\breve{A}_1 \breve{A}_2
   \breve{C}^{2}_3(\vecX) \ ) \nonumber \\
  && \qquad\qquad\qquad -\breve{B}_{23}^2 \breve{C}_1^2(\vecX) -\breve{B}_{13}^2 \breve{C}_2^2(\vecX) -\breve{B}_{12}^2
   \breve{C}_3^2(\vecX) \nonumber \\
  &&\qquad +2( \ \breve{B}_{13}\breve{B}_{23}\breve{C}_1(\vecX)\breve{C}_2(\vecX)
   +\breve{B}_{12}\breve{B}_{13}\breve{C}_2(\vecX)\breve{C}_3(\vecX)
  +\breve{B}_{12}\breve{B}_{23}\breve{C}_1(\vecX)\breve{C}_3(\vecX) \ ) \nonumber \\
  &&\qquad\qquad +4( \ \breve{A}_1\breve{B}_{23}\breve{C}_2(\vecX)\breve{C}_3(\vecX)
   +\breve{A}_2\breve{B}_{13}\breve{C}_1(\vecX)\breve{C}_3(\vecX)
   +\breve{A}_3\breve{B}_{12}\breve{C}_1(\vecX)\breve{C}_2(\vecX) \ )
   \nonumber \\
 && \
\end{eqnarray}
 . We introduce new complex coefficients


\begin{eqnarray}
&& (4\breve{A}_2\breve{A}_3-\breve{B}_{23}^2) \equiv \lambda_1 \ , (4\breve{A}_1\breve{A}_3-\breve{B}_{13}^2) \equiv \lambda_2 \ , (4\breve{A}_1\breve{A}_2-\breve{B}_{12}^2) \equiv \lambda_3  \nonumber \\ 
&&  (2\breve{B}_{13}\breve{B}_{23}+4\breve{A}_3\breve{B}_{12}) \equiv
  \mu_{12} \ ,  (2\breve{B}_{12}\breve{B}_{23}+4\breve{A}_2\breve{B}_{13}) \equiv
  \mu_{13} \ , \nonumber \\
&& (2\breve{B}_{12}\breve{B}_{13}+4\breve{A}_1\breve{B}_{23}) \equiv
  \mu_{23} \ 
\end{eqnarray}

 , then we can rewrite \(\Phi^{(k)}\) as follows.
\begin{equation}
  \Phi^{(k)}(\vecX,t) = \sum_{i=1}^3 \lambda_i \breve{C}_i^2(\vecX) \ + \ \sum_{
   \stackrel{ \scriptstyle (i,j)=(1,2) \mbox{ \small or}}{ \scriptstyle
   (2,3) \mbox{ \small or } (3,1)} } \mu_{ij} \breve{C}_i(\vecX) \breve{C}_j(\vecX) \qquad\qquad\quad
\end{equation}
 . Here, we write the real part and the imaginary part of \(\lambda_i\)
 and \(\mu_{ij}\). 
\begin{eqnarray}
  && \lambda_i \equiv \Re e \lambda_i + i \ \Im m \lambda_i \\[5pt]
      && \quad \Re e \lambda_i = -\frac{1}{\hbar^2} \biggl\{ \ m_1 \omega_1
	\cot[\omega_1(t-t_0)] \ m_2 \omega_2 \cot[\omega_2(t-t_0)] \
	(a_jb_k-a_kb_j)^2  \biggr. \nonumber \\
    && \qquad\qquad\qquad\qquad\qquad + m_2 \omega_2 \cot[\omega_2(t-t_0)] \ \frac{m_3}{(t-t_0)} \
	(b_jc_k-b_kc_j)^2 \nonumber \\
    && \qquad\qquad\qquad\qquad \biggl. + \frac{m_3}{(t-t_0)} \ m_1 \omega_1 \cot[\omega_1(t-t_0)] \
	(c_ja_k-c_ka_j)^2 \ \biggr\} +\frac{1}{4 \sigma_j^2 \sigma_k^2} \\[5pt]
      && \quad \Im m \lambda_i = -\frac{1}{2\hbar} \biggl\{ \ m_1 \omega_1 \cot[\omega_1(t-t_0)]\left( \frac{a_j^2}{\sigma_k^2}+\frac{a_k^2}{\sigma_j^2} \right) \nonumber \\
      &&  \qquad\qquad\qquad\qquad\quad +m_2
       \omega_2\cot[\omega_2(t-t_0)]\left(
				     \frac{b_j^2}{\sigma_k^2}+\frac{b_k^2}{\sigma_j^2} \right)  +\frac{m_3}{(t-t_0)}\left( \frac{c_j^2}{\sigma_k^2}+\frac{c_k^2}{\sigma_j^2}  \right) \ \biggr\} \nonumber \\
     && \
\end{eqnarray}
 and 
\begin{eqnarray}
 && \mu_{ij} \equiv \Re e \mu_{ij} +i \ \Im m \mu_{ij} \\[5pt]
  && \  \Re e \mu_{ij} = -\frac{2}{\hbar^2} \biggl\{
  m_1\omega_1\cot[\omega_1(t-t_0)] \ m_2\omega_2\cot[\omega_2(t-t_0)] \
  (a_jb_k-a_kb_j)(a_kb_i-a_ib_k) \biggr. \nonumber \\
  && \qquad\qquad\qquad\qquad \ +m_2 \omega_2 \cot[\omega_2(t-t_0)] \
   \frac{m_3}{(t-t_0)} \ (b_jc_k-b_kc_j)(b_kc_i-b_ic_k) \nonumber \\
  && \qquad\qquad\qquad\qquad\quad \  \biggl. + \frac{m_3}{(t-t_0)} \ m_1
   \omega_1 \cot[\omega_1(t-t_0)] \ (c_ja_k-c_ka_j)(c_ka_i-c_ia_k) \
   \biggr\} \nonumber \\
  && \ \\
  && \ \Im m \mu_{ij} = \frac{1}{\sigma_k^2 \hbar} \biggl\{
  m_1\omega_1\cot[\omega_1(t-t_0)]a_ia_j \biggr. \nonumber \\
  && \qquad\qquad\qquad\qquad\qquad\qquad\qquad\quad \
   \biggl. +m_2\omega_2\cot[\omega_2(t-t_0)]b_ib_j \
   +\frac{m_3}{(t-t_0)}c_ic_j \ \biggr\} \nonumber \\
  && \ \\
  && \qquad\qquad\qquad\qquad \bigl( \  (i,j,k) = (1,2,3)
  \ \mbox{or} \ (2,3,1) \ \mbox{or} \ (3,1,2) \ \bigr)\nonumber
\end{eqnarray}
 . Besides, we introduce new complex factors (\( La_{11}-Mu^{(k)}_0 \)).
\begin{eqnarray}
  && \quad La_{dd} = \sum_{i=1}^3 \lambda_i \alpha_d^{(i)2} ,\quad
  Mu_{dd} = \sum_{ \stackrel{ \scriptstyle (i,j)=(1,2) \mbox{ \small or}}{
   \scriptstyle (2,3) \mbox{ \small or } (3,1)} } \mu_{ij}
   \alpha_d^{(i)} \alpha_d^{(j)} \quad ( \ d=1,2,3 \ ) \nonumber \\
  && La_{df} = 2 \sum_{i=1}^3 \lambda_i \alpha_d^{(i)} \alpha_f^{(i)} , \quad
  Mu_{df} = \sum_{(i,j)} \mu_{ij}
   \left( \alpha_d^{(i)} \alpha_f^{(j)} + \alpha_d^{(j)} \alpha_f^{(i)}
   \right) \\
  && \qquad\qquad\qquad \bigl( \ (d,f)=(1,2) \ \mbox{or} \ (2,3) \
   \mbox{or} \ (3,1)  \ \bigr) \nonumber \\
  && La_d^{(k)} = \sum_{i=1}^3 \lambda_i \frac{d_i^{(k)}}{\sigma_i^2}
  \alpha_d^{(i)} , \quad 
  Mu_d^{(k)} = 0.5  \sum_{(i,j)} \mu_{ij} 
   \left( \frac{d_i^{(k)}}{\sigma_i^2} \alpha_d^{(j)} +
    \frac{d_j^{(k)}}{\sigma_j^2} \alpha_d^{(i)}  \right) \nonumber \\
  && \quad La_0^{(k)} = 0.25 \sum_{i=1}^3 \lambda_i \frac{d_i^{(k)2}}{\sigma_i^4},
  \quad 
  Mu_0^{(k)} = 0.25 \sum_{(i,j)} \mu_{ij}
   \frac{d_i^{(k)}}{\sigma_i^2} \frac{d_j^{(k)}}{\sigma_j^2}
\end{eqnarray}
 These \(\alpha_d^{(i)}\) are real,
\begin{eqnarray}
 && \quad C_i(\vecX) = - \alpha_1^{(i)} x_1 -\alpha_2^{(i)} x_2 -\alpha_3^{(i)}
	 x_3  \\[5pt]
 && \alpha_d^{(i)} \equiv 
	\left(\frac{m_1\omega_1}{\sin[\omega_1(t-t_0)]}a_i\right)a_d
	+\left(\frac{m_2\omega_2}{\sin[\omega_2(t-t_0)]}b_i\right)b_d
        +\left(\frac{m_3}{(t-t_0)}c_i\right)c_d \nonumber \\
 && \
\end{eqnarray}
 therefore the real and the imaginary parts of ( \(La_{11}-Mu^{(k)}_0\) )
 simply correspond to each of ( \(\lambda_{i},\mu_{ij}\) ), that is \\
\begin{equation}
 \quad \Re e La_{dd} = \sum_{i=1}^3 \Re e \lambda_i \alpha_d^{(i)2}
  \quad \mbox{etc..}
\end{equation}
 , then the real and imaginary parts of \(\Phi^{(k)}\) are 
\begin{equation}
  \Phi^{(k)}(\vecX,t) \equiv \Re e \Phi^{(k)}(\vecX,t) +i \ \Im m \Phi^{(k)}(\vecX,t) 
\end{equation}
 . With the help of the formulae above, we get
\begin{eqnarray}
  && \quad \Re e \Phi^{(k)}(\vecX,t) = -\frac{1}{\hbar^2}(\Re e
   La_{11}+\Re e Mu_{11})x_1^2  -\frac{1}{\hbar^2}(\Re e La_{22}+\Re e
   Mu_{22})x_2^2 \nonumber \\
  && \qquad\qquad\qquad\qquad \  -\frac{1}{\hbar^2}(\Re e La_{33}+\Re e Mu_{33})x_3^2
    -\frac{1}{\hbar^2}(\Re e  La_{12}+\Re e Mu_{12})x_1 x_2 \nonumber \\
  && \qquad\qquad\qquad\qquad  -\frac{1}{\hbar^2}(\Re e  La_{23}+\Re e Mu_{23})x_2 x_3 
   -\frac{1}{\hbar^2}(\Re e  La_{31}+\Re e Mu_{31})x_3 x_1  \nonumber \\
  && \qquad\qquad\qquad\qquad \  +\frac{1}{\hbar}(\Im m La^{(k)}_1 +\Im m
   Mu^{(k)}_1 ) x_1 +\frac{1}{\hbar}(\Im m La^{(k)}_2 +\Im m
   Mu^{(k)}_2 ) x_2 \nonumber \\
  && \qquad\qquad\qquad\qquad\quad +\frac{1}{\hbar}(\Im m La^{(k)}_3 +\Im m
   Mu^{(k)}_3 ) x_3 + \Re e La^{(k)}_0 +\Re e Mu^{(k)}_0 \\ [5pt]
  && \quad \Im m \Phi^{(k)}(\vecX,t) = -\frac{1}{\hbar^2}(\Im m
   La_{11}+\Im m Mu_{11})x_1^2  -\frac{1}{\hbar^2}(\Im m La_{22}+\Im m
   Mu_{22})x_2^2 \nonumber \\
  && \qquad\qquad\qquad\qquad \ \ -\frac{1}{\hbar^2}(\Im m La_{33}+\Im m Mu_{33})x_3^2
    -\frac{1}{\hbar^2}(\Im m  La_{12}+\Im m Mu_{12})x_1 x_2 \nonumber \\
  && \qquad\qquad\qquad\qquad  -\frac{1}{\hbar^2}(\Im m  La_{23}+\Im m Mu_{23})x_2 x_3 
   -\frac{1}{\hbar^2}(\Im m  La_{31}+\Im m Mu_{31})x_3 x_1  \nonumber \\
  && \qquad\qquad\qquad\qquad \ -\frac{1}{\hbar}(\Re e La^{(k)}_1 +\Re e
   Mu^{(k)}_1 ) x_1 -\frac{1}{\hbar}(\Re e La^{(k)}_2 +\Re e
   Mu^{(k)}_2 ) x_2 \nonumber \\
  && \qquad\qquad\qquad\qquad\quad -\frac{1}{\hbar}(\Re e La^{(k)}_3 +\Re e
   Mu^{(k)}_3 ) x_3 + \Im m La^{(k)}_0 +\Im m Mu^{(k)}_0
\end{eqnarray}
 . And another formula \( \breve{D}^{(k)}(\vecX) \) is
\begin{eqnarray}
  && \breve{D}^{(k)}(\vecX) \equiv \Re e \breve{D}^{(k)} + i \ \Im m
   \breve{D}(\vecX) \\[5pt]
  && \qquad \Re e \breve{D}^{(k)} =
   -\frac{d_1^{(k)2}}{4\sigma_1^2}-\frac{d_2^{(k)2}}{4\sigma_2^2}-\frac{d_3^{(k)2}}{4\sigma_3^2}
   \\
  && \qquad \Im m \breve{D}(\vecX) = D(\vecX)/\hbar
\end{eqnarray}
 . Next, the wave function is from (\ref{eqn053}), 
\begin{eqnarray}
  && \psi^{(k)}(\vecX,t) \propto \sqrt{ \frac{\pi^3\Delta^{*} }{
   |\Delta|^2} } \ \exp \left[ \ \frac{\Delta^{*}}{16|\Delta|^2}
		       \Phi^{(k)}(\vecX,t) + \breve{D}^{(k)}(\vecX,t)
			\ \right] \\[5pt]
  && \qquad\qquad\quad \equiv \ \breve{Q} \ \exp [ \ \Theta^{(k)}(\vecX,t) \ ]
\end{eqnarray}
 , where 
\begin{eqnarray}
  &&  \qquad\qquad\qquad\qquad\qquad \breve{Q} \equiv \Re e \breve{Q} + i \ \Im m \breve{Q} \\[5pt]
   &&  \Re e \breve{Q} = \sqrt{\frac{\pi^3}{|\Delta|}} \
       \cos\frac{\phi}{2} \ , \quad \Im m \breve{Q} =
       \sqrt{\frac{\pi^3}{|\Delta|}} \ \sin\frac{\phi}{2} \ \quad :
       \phi = \arctan\left( \Im m \Delta/ \Re e \Delta \right) \nonumber
       \\
   && \
\end{eqnarray}
 and 
\begin{eqnarray}
  && \qquad\qquad\qquad \Theta^{(k)}(\vecX,t) \equiv \Re e \Theta^{(k)}(\vecX,t) +i \ \Im m \Theta^{(k)}(\vecX,t)
   \\[5pt]  && \Re e \Theta^{(k)}(\vecX,t) = \frac{1}{16|\Delta|^2} ( \ \Re e \Delta \cdot \Re e \Phi^{(k)} +\Im m \Delta \cdot \Im m \Phi^{(k)} \ )
   + \Re e \breve{D}^{(k)} \\
  && \Im m \Theta^{(k)}(\vecX,t) = \frac{1}{16|\Delta|^2} ( \ \Re e
   \Delta \cdot \Im m \Phi^{(k)} -\Im m \Delta \cdot \Re e \Phi^{(k)} \ )
   + \Im m \breve{D}(\vecX) \nonumber \\
  &&
\end{eqnarray}
. Then, the real part and the imaginary part of the wave functions for respective \( k \) are as
 follows.
\begin{eqnarray}
 && \quad\qquad\quad \ \psi^{(k)}(\vecX,t) \equiv \Re e \psi^{(k)}(\vecX,t) +i \ \Im m
  \psi^{(k)}(\vecX,t) \\[5pt]
 && \Re e \psi^{(k)}(\vecX,t) \propto \exp[ \ \Re e \Theta^{(k)} \ ]
 \left( \ \Re e \breve{Q} \cdot \cos[\Im m \Theta^{(k)}] -\Im m \breve{Q}
  \cdot \sin[\Im m \Theta^{(k)}] \ \right) \nonumber \\
 && \qquad\qquad\qquad = \sqrt{\frac{\pi^3}{|\Delta|}} \ \exp[ \ \Re e
  \Theta^{(k)} \ ]\cdot\cos\left[ \ \Im m \Theta^{(k)} + \frac{\phi}{2} \ \right] \\
 && \Im m \psi^{(k)}(\vecX,t) \propto \exp[ \ \Re e \Theta^{(k)} \ ]
 \left( \ \Re e \breve{Q} \cdot \sin[\Im m \Theta^{(k)}] +\Im m \breve{Q}
  \cdot \cos[\Im m \Theta^{(k)}] \ \right) \nonumber \\
 && \qquad\qquad\qquad = \sqrt{\frac{\pi^3}{|\Delta|}} \ \exp[ \ \Re e
 \Theta^{(k)} \ ]\cdot\sin\left[ \ \Im m \Theta^{(k)} + \frac{\phi}{2} \ \right]
\end{eqnarray}
 Therefore the total wave function summed with (\( k=0-7 \)) is 
\begin{eqnarray}
  && \psi^{(total)}(\vecX,t) \equiv \Re e \psi^{(total)}(\vecX,t) +i \  \Im
   m \psi^{(total)}(\vecX,t) \\[5pt]
  && \qquad \Re e \psi^{(total)} = C \sqrt{\frac{\pi^3}{|\Delta|}} \ \sum_{k=0}^7 \exp[ \ \Re e
  \Theta^{(k)} \ ]\cdot\cos\left[ \ \Im m \Theta^{(k)} + \frac{\phi}{2}
			    \ \right] \\
  && \qquad \Im m \psi^{(total)} = C \sqrt{\frac{\pi^3}{|\Delta|}} \ \sum_{k=0}^7 \exp[ \ \Re e
  \Theta^{(k)} \ ]\cdot\sin\left[ \ \Im m \Theta^{(k)} + \frac{\phi}{2} \ \right]
\end{eqnarray}
 , where we introduced a real normalization constant C.  Then the quantum
 mechanical probability density function of whole system becomes as follows.
\begin{eqnarray}
  && \rho^{(total)}(\vecX,t) = \Re e^2 \psi^{(total)} + \Im m^2
   \psi^{(total)} \nonumber \\[5pt]
  && \qquad   = C^2 \frac{\pi^3}{|\Delta|} \ \sum_{k=0}^7 \sum_{l=0}^7
		  \exp[ \ \Re e \Theta^{(k)} +\Re e \Theta^{(l)} \
		  ]\cdot\cos[ \ \Im m \Theta^{(k)} -\Im m
			     \Theta^{(l)} \ ]  \nonumber \\
  && \qquad = C^2 \frac{\pi^3}{|\Delta|} \biggl( \sum_{k=0}^7 
		  \exp[ \ 2\Re e \Theta^{(k)}\ ]  \nonumber \\
  && \qquad\qquad\qquad\qquad +2\sum_{k<l}^{0-7} \exp [ \ \Re e
   \Theta^{(k)}+ \Re e \Theta^{(l)} \ ]\cdot\cos[ \ \Im m
   \Theta^{(k)}-\Im m \Theta^{(l)} \ ] \biggr) \label{eqn097} \nonumber
   \\
  &&
\end{eqnarray}
 The first term in \(( \cdots )\) of equation (\(\ref{eqn097}\)) means
 eight wave packets which originally are the Gaussian packets at
 the initial time \(t_0\). So we may call them ``definitive'' part of the density
 functions for whole system, or simply ``packets''. While the second term, is their
 ``interference'' part. But I should emphasize that it should not be vanished by quantum decoherence here! In fact, the ''interference'' part is different from what we have to observe. 

 Because now we are not interested in the information about whole system which contains three particles. We are only interested in the information about a sub-system, particle-1. So we have to average out the information about particle-2 and particle-3. Then we can get the information
 about particle-1 only, that is, the reduced density function for particle-1.

\begin{equation}
  \tilde{\rho}_1^{(reduced)}(x_1,t) \equiv \int_{-\infty}^{\infty}
   \int_{-\infty}^{\infty}dx_2 \ dx_3 \ \rho^{(total)}(\vecX,t) \label{eqn098}
\end{equation}
We substitute (\(\ref{eqn097}\)) into (\(\ref{eqn098}\)),
\begin{eqnarray}
  && \tilde{\rho}_1^{(reduced)}(x_1,t) = C^2 \frac{\pi^3}{|\Delta|}
   \biggl( \sum_{k=0}^7 \int_{-\infty}^{\infty}
   \int_{-\infty}^{\infty}dx_2 \ dx_3 \ \exp[ \ 2\Re e \Theta^{(k)}\ ]  \nonumber \\
  && \qquad +2\sum_{k<l}^{0-7} \int_{-\infty}^{\infty}
   \int_{-\infty}^{\infty}dx_2 \ dx_3 \ \exp [ \ \Re e
   \Theta^{(k)}+ \Re e \Theta^{(l)} \ ]\cdot\cos[ \ \Im m
   \Theta^{(k)}-\Im m \Theta^{(l)} \ ] \biggr) \nonumber \\
  && \ \label{eqn098b}
\end{eqnarray}

Here we notice that there are two kinds of ``the interference''. When there is
a interference between different packets, macroscopic states of
particle-1 which is included in each packet may be the same. For understanding this, it is simple that we think of the
initial states. 

From eq.(\(\ref{eqn052}\)),
\begin{equation}
  \begin{array}{cl}
    k=0,2,3,6 & \mbox{: \ Packets around (\(x_1=0\)) at initial time (\(t=t_0\)).} \\
    k=1,4,5,7 & \mbox{: \ Packets around (\(x_1=d_1\)) at initial time (\(t=t_0\)). }
  \end{array}
\end{equation}

The 8 packets in the (\(x_1,x_2,x_3\)) space are classified into these
two groups. Interferences between packets in the same group mean the transitions between states of particle-2 or of particle-3, not of particle-1. 
 After the transitions, particle-1 remains in the same macroscopic state yet. Therefore they are not the true interference between different macroscopic states of particle-1 which we really want to see.  

Now we have to classify these 8 packets into two groups above, and we have to add these interference terms between packets in the same group to "definitive" parts. Then we will get new packets. We should regard them as the effective macroscopic states for the particle-1. This procedure was suggested from numerical simulations. Because interferences between packets in same group also grow into new packet. Finally we can get as follows. \\[5pt]
The packet which was around (\(x_1=0\)) at initial time \(t_0\).：
\begin{eqnarray}
  && \tilde{\rho}_{1\_0}^{\mbox{ \tiny eff}}(x_1,t) \equiv C^2 \frac{\pi^3}{|\Delta|}
    \sum_{k=0,2,3,6} \biggl( \int_{-\infty}^{\infty}
   \int_{-\infty}^{\infty}dx_2 \ dx_3 \ \exp[ \ 2\Re e \Theta^{(k)}\ ]
   \nonumber  \\
  && \quad +2\sum_{l=0,2,3,6}^{k<l} \int_{-\infty}^{\infty}
   \int_{-\infty}^{\infty}dx_2 \ dx_3 \ \exp [ \ \Re e
   \Theta^{(k)}+ \Re e \Theta^{(l)} \ ]\cdot\cos[ \ \Im m
   \Theta^{(k)}-\Im m \Theta^{(l)} \ ] \biggr) \nonumber \\
  && \ \label{eqn_effrho01}
\end{eqnarray}
The packet which was around (\(x_1=d_1\)) at initial time \(t_0\).：
\begin{eqnarray}
  && \tilde{\rho}_{1\_d}^{\mbox{ \tiny eff}}(x_1,t) \equiv C^2 \frac{\pi^3}{|\Delta|}
    \sum_{k=1,4,5,7} \biggl( \int_{-\infty}^{\infty}
   \int_{-\infty}^{\infty}dx_2 \ dx_3 \ \exp[ \ 2\Re e \Theta^{(k)}\ ]
   \nonumber  \\
  && \quad +2\sum_{l=1,4,5,7}^{k<l} \int_{-\infty}^{\infty}
   \int_{-\infty}^{\infty}dx_2 \ dx_3 \ \exp [ \ \Re e
   \Theta^{(k)}+ \Re e \Theta^{(l)} \ ]\cdot\cos[ \ \Im m
   \Theta^{(k)}-\Im m \Theta^{(l)} \ ] \biggr) \nonumber \\
  && \ \label{eqn_effrhod1}
\end{eqnarray}
Their effective interference term.：

\begin{eqnarray}
 &&  \tilde{\rho}_{1\_int}^{\mbox{ \tiny eff}}(x_1,t) \equiv \nonumber \\
 && \quad  4C^2
   \frac{\pi^3}{|\Delta|} \sum_{  \stackrel{ \scriptstyle k=0,2,3,6}{
    \scriptstyle l=1,4,5,7} }^{k<l} \int_{-\infty}^{\infty}
   \int_{-\infty}^{\infty}dx_2 \ dx_3 \ \exp [ \ \Re e
   \Theta^{(k)}+ \Re e \Theta^{(l)} \ ]\cdot\cos[ \ \Im m
   \Theta^{(k)}-\Im m \Theta^{(l)} \ ] \nonumber \\ 
 && \quad  \label{eqn_effinterf}
\end{eqnarray}

Finally, we get the reduced density for particle 1 as follow.
\begin{equation}
  \tilde{\rho}_1(x_1,t) = \tilde{\rho}_{1\_0}^{\mbox{ \tiny
   eff}}(x_1,t) + \tilde{\rho}_{1\_d}^{\mbox{ \tiny eff}}(x_1,t) + \tilde{\rho}_{1\_int}^{\mbox{ \tiny eff}}(x_1,t)
\end{equation}

Numerical calculations are used for integrations in eq.(\ref{eqn098}),(\ref{eqn098b}),(\ref{eqn_effrho01})-(\ref{eqn_effinterf}), and
final normalization which determines constant \(C^2 \frac{\pi^3}{|\Delta|}\).


\begin{thebibliography}{9}

\bibitem{1}
 W.H.Zurek, Los Alamos Science \textbf{27},2 (2002).

\bibitem{2}
 S.Takagi, {\em Kyoshiteki Tonneru Gensho} (Macroscopic tunneling)(IwanamiShoten,Tokyo,1997)[in Japanese].

\bibitem{3}
 H.Everett, Reviews of Modern Physics \textbf{29}, 454–462 (1957).

\bibitem{4}
 T.Ishikawa, in {\it  Proceedings of the 12th Asia Pacific Physics Conference (APPC12),Japan,2013} edited by M. Sasao, JPS Conf. Proc.\textbf{1}, 012133 (2014).

\bibitem{5}
 A.O.Caldeira and A.J.Leggett, Physical Review A \textbf{31},1059 (1985).

\bibitem{6}
 A.O.Caldeira and A.J.Leggett, Physica A \textbf{121}, 587 (1983).

\bibitem{7}
 I.Ragnarsson and S.G.Nilsson, {\em Shapes and Shells in Nuclear Structure }(Cambridge University Press,2005).

\bibitem{8}
 H.Kuratsuji, {\em Genshikaku Kenkyu} \textbf{28} No.2, 3-33 (Genshikaku Danwakai,1983).

\end{thebibliography}
\end{document}